\documentclass{aa} 
\usepackage{graphicx}
\usepackage{txfonts}
\usepackage{lipsum}
\usepackage{subcaption}                                      
\usepackage{lscape}             
\usepackage{placeins}           
\usepackage{amsmath}
\usepackage{hyperref}

\begin{document}

   \title{\textit{EP-FXT} Observations of Abell 1795: X-ray properties and structure out to \( R_{200} \)}

\author{
  {Y. Tang}\inst{1,2}%
  \and {S.M. Jia}\inst{1}\corrauth{jiasm@ihep.ac.cn}%
  \and {H.H. Zhao}\inst{2}\corrauth{zhaohh@njnu.edu.cn}%
  \and {C.K. Li}\inst{1}\email{lick@ihep.ac.cn}%
  \and {H. Yu}\inst{3}\email{yuheng@bnu.edu.cn}%
  \and {Y. Chen}\inst{1}\email{ychen@ihep.ac.cn}%
  \and {S.S. Weng}\inst{2}\email{wengss@njnu.edu.cn}%
  \and {X.Y. Zheng}\inst{1,3}\email{xyzheng@ihep.ac.cn}%
  \and {\\H. Feng}\inst{1}\email{hfeng@ihep.ac.cn}%
  \and {L.M. Song}\inst{1}\email{songlm@ihep.ac.cn}%
  \and {C.Z. Liu}\inst{1}\email{liucz@ihep.ac.cn}%
  \and {F.J. Lu}\inst{1}\email{lufj@ihep.ac.cn}%
  \and {S.N. Zhang}\inst{1}\email{zhangsn@ihep.ac.cn}%
  \and {W.M. Yuan}\inst{4}\email{wmy@nao.cas.cn}%
  \and {S. Andreon}\inst{8}\email{stefano.andreon@inaf.it}%
  \and {\\J.F. Wang}\inst{6}\email{jfwang@xmu.edu.cn}%
  \and {W.W. Cui}\inst{1}\email{cuiww@ihep.ac.cn}%
  \and {J. Guan}\inst{1}\email{jguan@ihep.ac.cn}%
  \and {C.C. Jin}\inst{4,5,7}\email{ccjin@nao.cas.cn}%
  \and {Y. Liu}\inst{4}\email{liuyuan@bao.ac.cn}%
  \and {J. Zhang}\inst{1}\email{zhangjuan@ihep.ac.cn}%
  \and {H.S. Zhao}\inst{1}\email{zhaohs@ihep.ac.cn}%
  \and {X.F. Zhao}\inst{1}\email{zhaoxf@ihep.ac.cn}%
  }

\institute{%
    State Key Laboratory of Particle Astrophysics, Institute of High Energy Physics, Chinese Academy of Sciences, Beijing 100049, China \\
  \email{jiasm@ihep.ac.cn}
  \and School of Physics and Technology, Nanjing Normal University, Nanjing, 210023, Jiangsu, China \\
  \email{zhaohh@njnu.edu.cn}
  \and School of Physics and Astronomy, Beijing Normal University, Beijing 100875, China
  \and National Astronomical Observatories, Chinese Academy of Sciences, 20A Datun Road, Beijing 100101, China
  \and Institute for Frontiers in Astronomy and Astrophysics, Beijing Normal University, Beijing 102206, China
  \and Department of Astronomy, Xiamen University, Xiamen 361005, China
  \and School of Astronomy and Space Science, University of Chinese Academy of Sciences, 19A Yuquan Road, Beijing 100049, China
  \and INAF–Osservatorio Astronomico di Brera, via Brera 28, 20121, Milano, Italy
}

   \date{Received September 30, 20XX}

  \abstract
  {We present deep X-ray observations of the nearby X-ray-luminous galaxy cluster Abell 1795 obtained with the \textit{Einstein Probe} Follow-up X-ray Telescope (\textit{EP-FXT}), with a total exposure time of $\sim 480$ ks. Exploiting the large field of view and low particle background of \textit{EP-FXT}, we directly measured the radial temperature profile of A1795 out to $R_{200}$ with full azimuthal coverage, which was shown to increase in line with the radius within $6\arcmin$ and then gradually decline toward larger radii. The surface-brightness residual map and 2D thermodynamic maps reveal a clockwise spiral structure extending from the cluster core towards the south-east, which traces low-temperature and low-entropy gas, which is also consistent with sloshing-induced cold fronts. In the north-west, the surface-brightness-enhanced region exhibits an arc-like high-temperature feature and both the temperature-derived and density-derived Mach numbers support the assumption that it is a weak shock. These substructures can be explained by a binary merger scenario: the perturbing subcluster induces sloshing during its first passage past the primary core and its subsequent return passage through the ICM might drive the shock towards the north-west. Our results indicate that relaxed galaxy clusters such as A1795 might still retain signatures of dynamical activity.}


   \maketitle

   \nolinenumbers

\section{Introduction}
Galaxy clusters form through the gravitational collapse of primordial density fluctuations and contain between hundreds and thousands of member galaxies, making them the largest self-gravitating systems in the Universe. The intracluster medium (ICM) is a hot, tenuous, ionised plasma that fills the entire cluster volume and serves as the primary source of its X-ray emission \citep{Kravtsov2012,Vikhlinin2014}. Galaxy clusters are typically classified as either relaxed or unrelaxed based on their dynamical state and structural symmetry. 

In their most relaxed state, galaxy clusters generally appear regular and nearly symmetric in X-ray emission. They are therefore regarded, based on the assumption of hydrostatic equilibrium, as ideal laboratories for precision cosmology. However, increasingly deep, high-quality observations have revealed that the formation and evolution of these apparently highly relaxed systems are, in fact, fundamentally dynamical: frequent mergers and continuous mass accretion induce significant dynamical effects in the ICM, driving the system away from strict hydrostatic and thermal equilibrium \citep{Markevitch2007,Kravtsov2012,Walker2017,Eckert2019,Rahaman2022}. Consequently, understanding and characterising these perturbations is a crucial step towards constraining the formation and evolution of galaxy clusters and reconstructing the assembly history of large-scale structure in the Universe.

Such gravitational perturbations, induced by mergers and mass accretion, are manifested in X-rays through a variety of structural features, including shocks and cold fronts. During cluster mergers, shocks are commonly generated. When an infalling subcluster moves supersonically through the ICM, it drives a shock front that compresses the gas and heats it via dissipation on short timescales, producing discontinuities in thermodynamic quantities, such as density and temperature, across the interface \citep{Markevitch2002,Markevitch2005,Russell2010}. Observationally, merger shocks are seen as sharp edges in the X-ray surface brightness \citep{Markevitch2007}. Typical examples of merger shocks include 1E~0657--56 (i.e., the Bullet Cluster; \citealt{Markevitch2002}) and A520 \citep{Markevitch2005}. In contrast, cold fronts also appear as sharp edges in the X-ray surface brightness, but the ICM on the inner side of the front is cooler and denser, while the gas outside is hotter; the pressure remains approximately continuous across the interface, whereas the entropy exhibits a clear discontinuity. Cold fronts and their associated large-scale spiral structures have been widely identified in many galaxy clusters. They generally originate from the sloshing of low-entropy gas in the cluster core within the gravitational potential well, typically triggered by minor, off-axis mergers or close passages of subclusters \citep{Ascasibar2006,Roediger2011,ZuHone2011,ZuHone2018,Watson2025}. Representative examples include the Perseus cluster \citep{Churazov2003,Walker2017}, the Virgo cluster \citep{Roediger2011,Zheng2025}, Abell~2029 (A2029) \citep{Paterno2013,Watson2025}, and Abell~119 \citep{Watson2023}, among others.

Abell~1795 (A1795) is a nearby rich galaxy cluster at a redshift of $z=0.0622$, covering a fairly large angular extent, with $R_{200}\simeq2~\mathrm{Mpc}$ ($\approx26'$; \citealt{Sanderson2003,Vikhlinin2006}). The cluster core is dominated by a giant cD galaxy, which coincides with the brightest cluster galaxy (BCG), PGC~049005 \citep{Ebeling1998,Xu1998}. A1795 exhibits bright X-ray emission in the 2--10~keV band, with a luminosity of $L_{\mathrm{X}}\simeq1.0\times10^{45}~\mathrm{erg\,s^{-1}}$ \citep{Xu1998} and an average temperature of $kT\simeq6.1~\mathrm{keV}$ \citep{Vikhlinin2006}. It is generally classified as a morphologically regular, relaxed cool-core cluster \citep{Buote1996,Kokotanekov2018}.

In terms of the thermal properties, the radial temperature profile of A1795 has been measured by multiple X-ray observatories; however, the reported results are not fully consistent. \citet{Tamura2001} used a single \textit{XMM-Newton} EPIC observation and detected the ICM temperature increases from 3.2~keV to 7.0~keV out to $8\arcmin$. Using \textit{Chandra} mosaic observations, \citet{Vikhlinin2006} found that the temperature peaks at 6.7~keV at $R\approx5\arcmin$, and then declines to 5.4~keV by $R\approx17\arcmin$, with an average temperature of 6.1~keV. Multi-pointing \textit{Suzaku} mosaic observations further extended the temperature profile along the northern and southern directions. In the north, the temperature peaks at 5.8~keV at $R\approx8\arcmin$ and declines to 2.1~keV by $R\approx26\arcmin$. In the south, the temperature peaks at 5.4~keV at $R\approx8\arcmin$ and decreases to 1.7~keV by $R\approx18\arcmin$ \citep{Bautz2009}. Moreover, the X-COP joint analysis, combining \textit{XMM-Newton} mosaic observations with \textit{Planck} Sunyaev--Zel'dovich constraints, reported a peak temperature of 5.6~keV at $R\approx4\arcmin$ and a decline to 2.7~keV by $R\approx27\arcmin$ \citep{Ghirardini2019}. Overall, in the central region, these measurements are broadly consistent within the uncertainties despite differences in the adopted spatial regions and instrumental coverage. In the outskirts, the \textit{Suzaku} and X-COP temperature profiles remain consistent with each other, but lie below the \textit{Chandra} profile. Therefore, to more comprehensively investigate the temperature profile of A1795, it is important to conduct X-ray observations with a larger field of view (FOV) and sufficient sensitivity.

\begin{table*}[t]
\caption{\textit{EP-FXT} observations of A1795 used in this paper.}
\label{tab:fxt_obslog}    
\centering               
\begin{tabular}{l c c c c c c} 
\hline\hline              
ObsID & RA (deg) & Dec (deg) & Date & Exposure time (ks) & Filter & Instrument \\  
\hline                      
11904194479 & 207.22 & 26.60 & 2024-06-11 & 79.33 & thin   & A\&B \\
13600006485 & 207.22 & 26.60 & 2024-06-26 & 87.78 & thin   & A\&B \\
11900041216 & 207.22 & 26.60 & 2025-01-10 & 21.85 & thin   & A\&B \\
11900047488 & 207.22 & 26.60 & 2025-01-14 & 42.81 & thin   & A\&B \\
11900054400 & 207.22 & 26.60 & 2025-01-19 & 31.01 & medium & A\&B \\
11900065280 & 207.22 & 26.60 & 2025-01-25 & 39.88 & medium & A\&B \\
11900157824 & 207.22 & 26.60 & 2025-03-28 & 47.64 & thin   & A\&B \\
13600172800 & 207.22 & 26.60 & 2025-04-04 & 38.95 & medium & A\&B \\
11900274944 & 207.22 & 26.60 & 2025-06-17 & 58.80 & medium & A\&B \\
11900285696 & 207.22 & 26.60 & 2025-06-26 & 32.70 & thin   & A\&B \\
\hline                                  
\end{tabular}
\end{table*}

In terms of the structural properties, joint spectral analyses of the \textit{Chandra}, \textit{XMM-Newton}, and \textit{Suzaku} data have shown that the core of A1795 hosts a typical multiphase cool core \citep{Gu2012,Xu1998}. Multiwavelength observations further revealed pronounced dynamical disturbances around the cool core. In the X-ray band, a cold front is detected to the south at $R \simeq 86\,\mathrm{kpc}$ and has been interpreted as having been induced by gas sloshing \citep{Markevitch2001}. At optical/UV wavelengths and in cold-gas tracers, an observed ``cooling wake'' that is spatially coincident with H$\alpha$/UV emission and cold molecular gas suggests that radiative cooling remains active on kiloparsec scales of several tens to hundreds \citep{Fabian2001,Ettori2002,Salom2003,McDonald2009,Ehlert2015}. The \textit{Chandra} temperature map shows clear inhomogeneities within the central $\sim 5\arcmin$, but it does not reveal a large-scale structure pattern owing to the limited FOV \citep{Ehlert2015}. On larger scales, \textit{Suzaku} observations using a one-direction mosaic traced the X-ray emission and temperature profile of A1795 out to nearly $R_{200}$. The results show that the temperature in the outskirts declines rapidly with radius, suggesting possible deviations from hydrostatic equilibrium around $R \sim R_{500}$ ($\approx 18'$) \citep{Bautz2009}. More recent joint analyses of \textit{Chandra} and \textit{Suzaku} data have shown that even after identifying and excising candidate gas clumps, the entropy profile of A1795 near $R_{200}$ still significantly deviates from the simple power-law form expected under pure gravitational collapse, suggesting complex gas physics in the cluster outskirts \citep{Kovacs2023}. Meanwhile, although weak gas clumps are detected, no prominent large-scale substructures are evident \citep{Kovacs2023}. Therefore, although A1795 appears globally regular in its projected X-ray morphology, it already exhibits some departures from hydrostatic equilibrium and complex thermodynamic structures from the innermost cool-core region out to nearly $R_{200}$. This motivates a more systematic and in-depth investigation of the gas distribution and dynamical state of A1795 using wide-field X-ray imaging observations.

Follow-up X-ray Telescope (\textit{EP-FXT}) is one of the key instruments onboard the \textit{Einstein Probe} (EP) mission, which was successfully launched on 9 January 2024 \citep{Yuan2022,Yuan2025}. \textit{EP-FXT} has a FOV of about $1~\mathrm{deg}^2$, a pixel scale of $9.6''$, and an on-axis angular resolution corresponding to a half-power diameter (HPD) of $22''$. It operates in the 0.3--10~keV energy band. Owing to its extremely low particle background \citep{Zhang2025}, \textit{EP-FXT} is particularly well suited for detecting large-scale, low-surface-brightness diffuse X-ray emission. 

In this paper, we present a systematic analysis of the X-ray properties of A1795 based on deep \textit{EP-FXT} imaging observations. This paper is organised as follows. In Sect.~\ref{sec:data} we describe the data sets used in our analysis and the X-ray data reduction. In Sect.~\ref{sec:analysis} we present the data analysis and results. In Sect.~\ref{sec:discussion} we discuss the implications of our findings. Our conclusions are summarised in Sect.~\ref{sec:conclusion}. Throughout this paper, we assume a standard $\Lambda$CDM cosmology with $H_0 = 70~\mathrm{km~s^{-1}~Mpc^{-1}}$, $\Omega_{\mathrm{m}} = 0.3$, and $\Omega_{\Lambda} = 0.7$. At a redshift of A1795 ($z = 0.0622$), the angular scale is $1\arcmin = 71.9~\mathrm{kpc}$. 

\begin{figure}[ht!]
\centering
\includegraphics[width=\hsize]{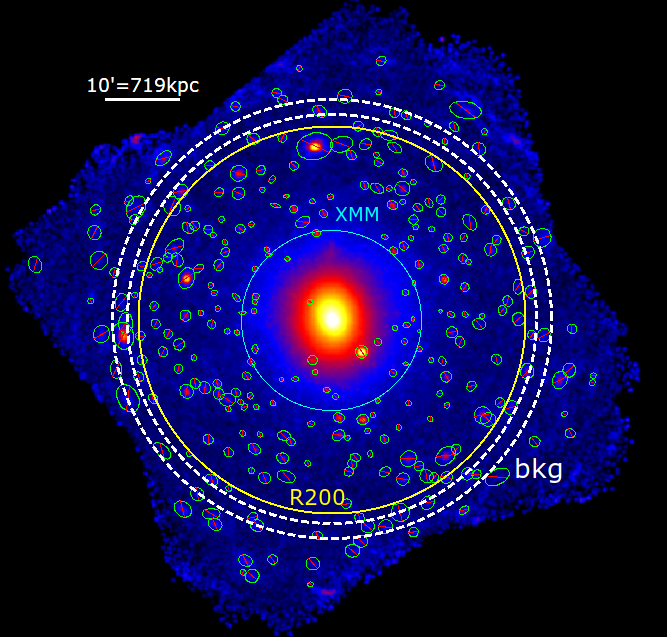}
\caption{Combined \textit{EP-FXT} X-ray image of A1795 in the 0.3--7.0~keV band. The image is vignetting-corrected, with the particle background subtracted, and then smoothed. White circles mark the background annulus at radii of $27.5\arcmin$--$29.5\arcmin$. Cyan circle indicates the approximate FOV of \textit{XMM-Newton}, and the yellow circle marks $R_{200}$ ($1.9~\mathrm{Mpc}$). Detected point sources are indicated by green circles.}
\label{fig:2}
\end{figure}

\section{Observations and data reduction}\label{sec:data}
As a calibration target for \textit{EP-FXT}, the galaxy cluster A1795 has been observed extensively. All data used in this work were acquired in full-frame (FF) mode with either the thin or medium optical filter, yielding a total good-time exposure of approximately $480$~ks across all observations (see Table~\ref{tab:fxt_obslog}).

We processed the data with \textit{EP-FXT} data analysis software (FXTDAS version 1.30; \citealt{Zhao2025})\footnote{\url{http://epfxt.ihep.ac.cn/analysis}}, together with general-purpose tools from CIAO (version 4.17)\footnote{\url{https://cxc.cfa.harvard.edu/ciao/}} and HEASoft (version 6.34)\footnote{\url{https://heasarc.gsfc.nasa.gov/docs/software/heasoft/}}. Calibrated event files were produced with \texttt{fxtchain}. To suppress out-of-time (OoT) events, we simulated OoT event lists for each observation using \texttt{fxtootest}, generated the corresponding OoT images, and subtracted them from the raw images with \texttt{ftimgcalc}. Since the particle background is not affected by vignetting, it was subtracted prior to exposure correction. We generated particle-background products for each observation using \texttt{fxtbkggen} \citep{Zhang2025} and subtracted them from the OoT-corrected images. The corresponding exposure maps were generated using \texttt{fxtexpogen} with \texttt{edgecovermask}=1. We then used CIAO's \texttt{reproject\_image} to register and merge the 0.3--7.0~keV OoT-corrected and particle-background-subtracted images, together with their exposure maps, in a common astrometric frame. The final vignetting-corrected image was obtained by dividing the merged counts image by the merged exposure map using \texttt{farith}. Point sources were detected on the image using \texttt{wavdetect} with scales of 2, 4, 8, and 16 pixels and the default \texttt{sigthresh}. Fig.~\ref{fig:2} shows the resulting 0.3--7.0~keV X-ray image, smoothed with a Gaussian kernel of \(\sigma=14.4\arcsec\) for display, with the detected source regions marked.

\section{Analysis and results}\label{sec:analysis}
\subsection{The surface brightness profile}

\begin{figure}[ht!]
\centering
\includegraphics[width=\hsize]{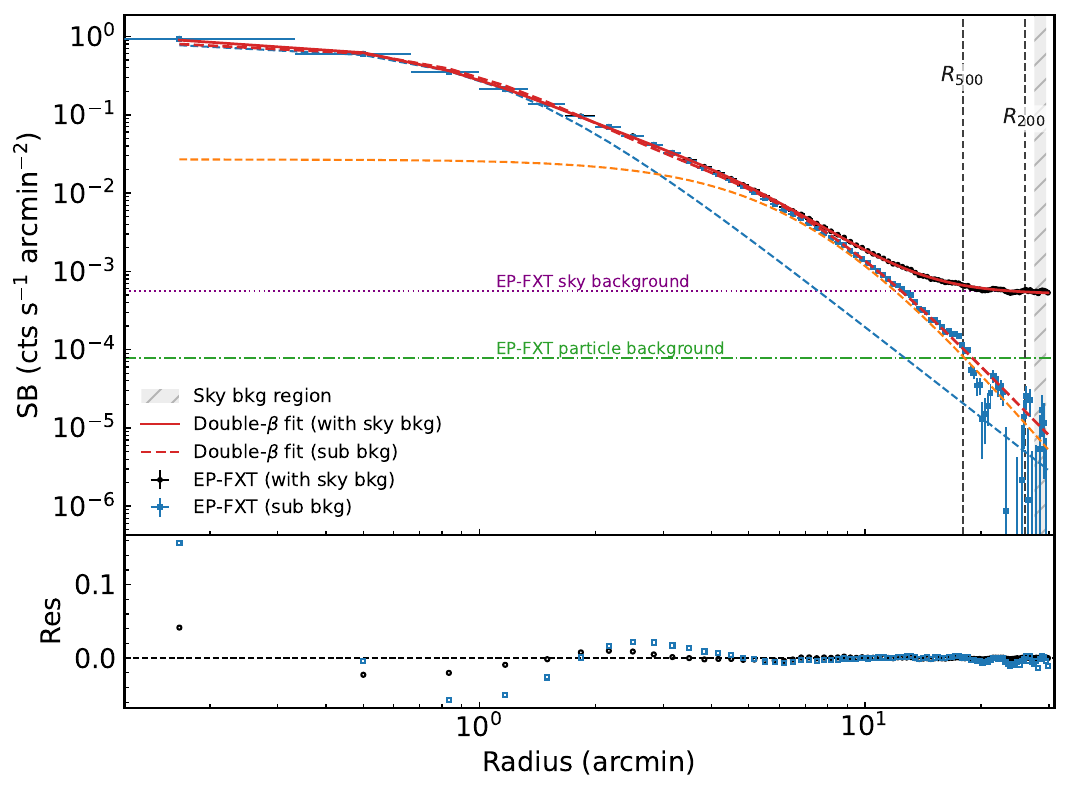}
\caption{\textit{EP-FXT} data after subtracting the particle background and correcting for vignetting, represented by the black points. The blue points represent the \textit{EP-FXT} surface-brightness data after subtracting both the particle and sky backgrounds. These data are fitted with a double-$\beta$ model without background terms, whose best-fit curve is shown as the red dashed line, while the orange and blue dashed lines indicate the two components of this model. The green and purple horizontal lines represent the \textit{EP-FXT} particle background from \texttt{fxtbkggen} and sky background levels, respectively.}
\label{fig:SB_2beta}
\end{figure}

To characterise the ICM distribution of A1795 observed with \textit{EP-FXT}, the detected point-source regions were excised from Fig.~\ref{fig:2} and inpainted with \texttt{dmfilth} using the \texttt{POISSON} method and a random seed of 0 to preserve the continuity of the diffuse surface brightness. We extracted the radial surface-brightness profile of A1795 in the 0.3--7.0~keV band using \texttt{pyproffit}\footnote{\url{https://pyproffit.readthedocs.io/en/latest/modules/pyproffit}} \citep{Eckert2020}, centred on the X-ray emission peak, as shown in Fig.~\ref{fig:SB_2beta}. The black points represent the \textit{EP-FXT} surface-brightness profile after subtracting the particle background and correcting for vignetting, extending out to approximately \(30\arcmin\). The profile becomes nearly flat in the outermost region, where the emission is dominated by the sky background. Therefore, we estimated the local sky-background level from the mean surface brightness in the \(27.5\arcmin\)--\(29.5\arcmin\) annulus outside \(R_{200}\), corresponding to the white annulus in Fig.~\ref{fig:2}. This level is shown by the purple horizontal line, \(\sim 5.6\times10^{-4}~\mathrm{cts\,s^{-1}\,arcmin^{-2}}\). For reference, the green horizontal line shows the \textit{EP-FXT} particle-background level, \(\sim 7.9\times10^{-5}~\mathrm{cts\,s^{-1}\,arcmin^{-2}}\), generated with \texttt{fxtbkggen} and already subtracted in Sect.~\ref{sec:data}. The blue points show the resulting ICM surface-brightness profile after subtracting both background components. We fit the profile with a double-$\beta$ model of the form
\begin{equation}
S(r) = A\left[1+\left(\frac{r}{r_{c1}}\right)^2\right]^{-3\beta_1+0.5}
     + B\left[1+\left(\frac{r}{r_{c2}}\right)^2\right]^{-3\beta_2+0.5}
     + C.
\label{eq:double_beta_A1795}
\end{equation}
The best-fit parameters are listed in Table~\ref{tab:doublebeta_params}. We also plot the two components of the double-$\beta$ model, shown as blue and orange dashed lines, which reveal the presence of a dense core in the central region of A1795, as already identified in \citet{Fabian1994}. 

Given that the ICM surface brightness in the outskirts is lower than the background level, we assessed the significance of the background-subtracted ICM emission. Owing to the long \textit{EP-FXT} exposure and reliable background characterisation, the ICM emission between \(R_{500}\) and \(R_{200}\) is still detected with a signal-to-noise ratio (S/N) of approximately 7.0, allowing us to trace the X-ray emission of A1795 out to \(R_{200}\). Therefore, \textit{EP-FXT} offers favourable conditions for studying large-scale diffuse X-ray emission.

\begin{table}[ht!]
\caption{Best-fitting parameters of the double-$\beta$ model with $1\sigma$ uncertainties.}
\label{tab:doublebeta_params}
\centering
\begin{tabular}{l c}
\hline\hline
Parameter      & Value \\
\hline
$\beta_{1}$        & $0.810 \pm 0.113$ \\
$r_{\mathrm{c}1}$  & $1.163 \pm 0.160$ \\
$A$                & $0.803 \pm 0.046$ \\
$\beta_{2}$        & $1.155 \pm 0.130$ \\
$r_{\mathrm{c}2}$  & $7.294 \pm 1.068$ \\
$B$                & $0.0269 \pm 0.0082$ \\
$C$                & $(0.0 \pm 3.8)\times10^{-6}$ \\
\hline
\end{tabular}
\tablefoot{Best-fit parameters of the double-$\beta$ model with $1\sigma$ uncertainties, based on the background-subtracted surface-brightness profile. $r_{\mathrm{c}1}$ and $r_{\mathrm{c}2}$ are given in units of arcmin. $A$, $B$, and $C$ are in units of cts\,s$^{-1}$\,arcmin$^{-2}$.}
\end{table}

\subsection{The radial temperature profile}\label{subsec:radial_temperature_profile}
\begin{figure}[ht!]
   \centering
   \includegraphics[width=\hsize]{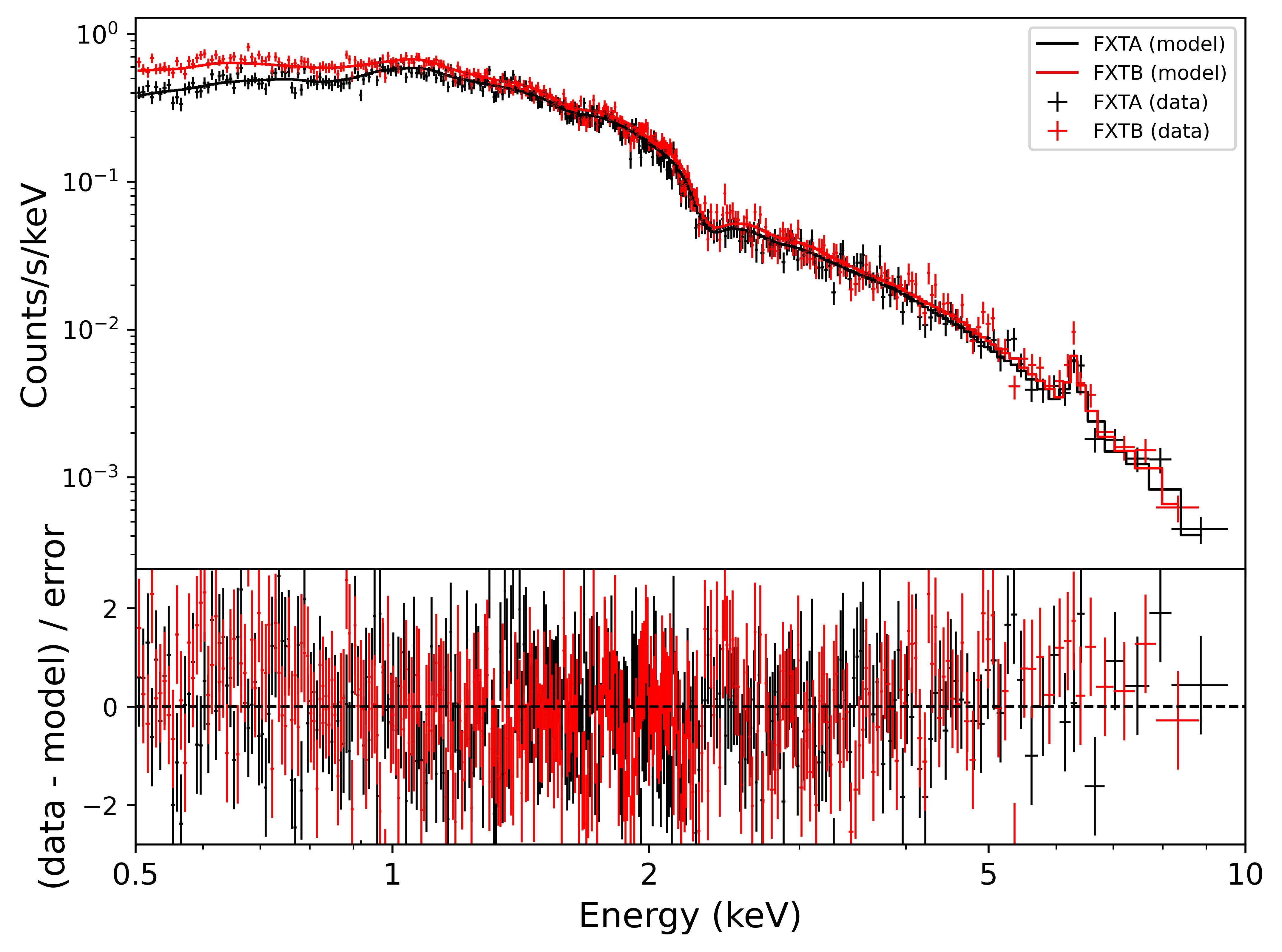}
   \caption{\textit{EP-FXT} X-ray spectrum extracted from the \(1.5\)--\(2.2\) arcmin annulus of A1795 (ObsID: 13600006485). Black and red data points represent the background-subtracted X-ray spectra obtained from the FXT-A and FXT-B modules, respectively. Solid lines indicate the best-fit models. Lower panel shows the residuals.}
     \label{fig:spec_show}
   \end{figure}

Based on the large FOV of \textit{EP-FXT}, we are able to efficiently explore the outer regions of A1795, up to \(R_{200}\). Centring on the X-ray peak, we used XSELECT v2.5b to extract spectra from 11 concentric annuli in the 0.5--10.0 keV energy band for FXT-A and FXT-B separately, with each annulus containing more than 80,000 photons. The ancillary response files (ARFs) are generated using the \textit{EP-FXT} response tool \texttt{fxtarfgen} with the \texttt{extend} parameter set to 1 to account for the extended nature of the source. The response matrix files (RMFs) were obtained using \texttt{fxtrmfgen}. The RMFs characterise the detector energy redistribution and spectral resolution. The spectra were grouped using \texttt{grppha} with a minimum of 30 counts per bin to ensure the applicability of the C-Statistic. The sky background was extracted from an annular region at radii of \(27.5\)--\(29.5'\) (see the white annulus in Fig.~\ref{fig:2}). The spectra were fitted using XSPEC v12.14.1 with a \texttt{TBABS*APEC} model, adopting APEC v3.0.9. We adopted the solar abundance table from \citet{AndersGrevesse1989}, with the redshift fixed at 0.0622 and the hydrogen column density fixed at $1.20 \times 10^{20}\,\mathrm{cm}^{-2}$ (\citealp{Willingale2013}). The temperature ($kT$), abundance ($Z$), and normalisation were left free. The best-fit parameters are summarised in Table~\ref{tab:a1795_abs1T_annuli}. For the outermost five annuli, where the metal abundance is poorly constrained, we also report the fitting results obtained with the abundance fixed at $Z=0.1\,Z_{\odot}$. Here, we present the background-subtracted spectra together with the best-fitting models for the third annulus extracted from FXT-A and FXT-B of observation ID 13600006485 (see Fig.~\ref{fig:spec_show}). 

\begin{figure}[ht!]
   \centering
   \includegraphics[width=\hsize]{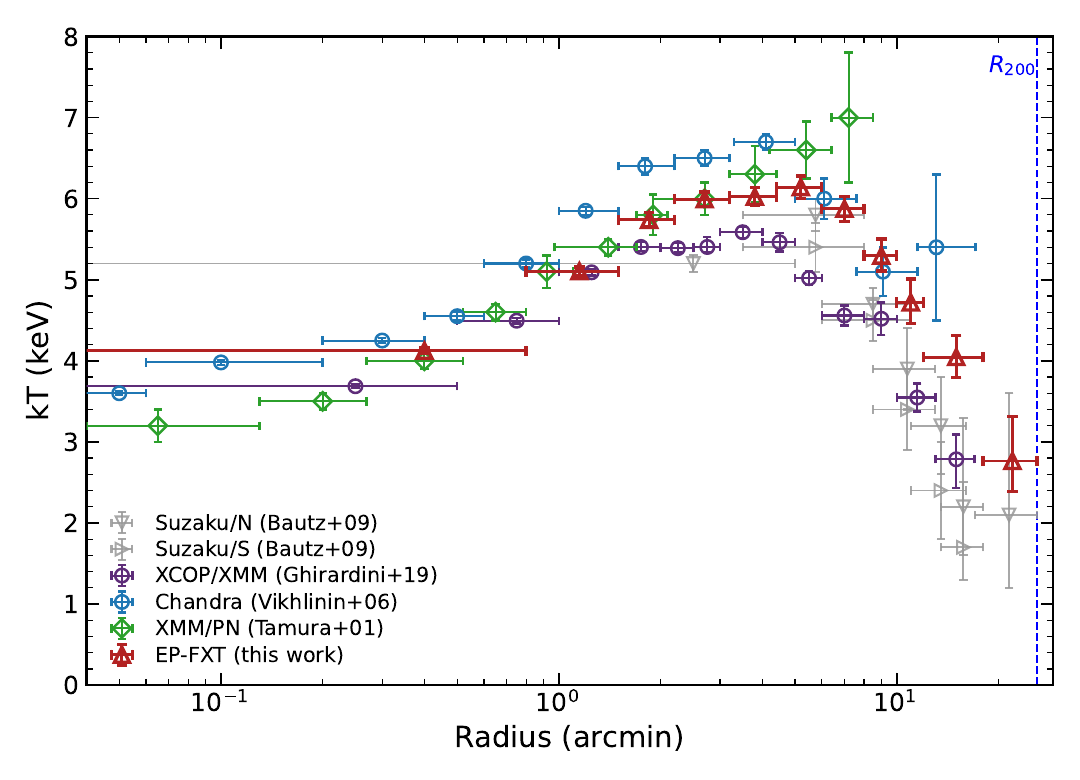}
   \caption{Projected radial temperature profiles of A1795. The symbols and colours indicate different measurements: \textit{EP-FXT} (red), \textit{XMM-Newton} \citep{Tamura2001} (green), \textit{Chandra} \citep{Vikhlinin2006} (blue), and X-COP (\textit{XMM-Newton}) \citep{Ghirardini2019} (purple). Grey inverted and right-pointing triangles denote the \textit{Suzaku} measurements in the northern and southern directions, respectively \citep{Bautz2009}.}
   \label{fig:kT_radial}
   \end{figure}

The projected radial temperature profile of A1795 directly measured with \textit{EP-FXT} is shown in Fig.~\ref{fig:kT_radial}. The temperature increases with radius in the inner region and peaks at $R\approx 6\arcmin$, then declines towards larger radii, with our measurements extending out to $R_{200}$ ($\approx26'$). We compare our measurements with previous \textit{XMM-Newton} results \citep{Tamura2001}, \textit{Chandra} measurements \citep{Vikhlinin2006}, the X-COP temperature profile from \textit{XMM-Newton} \citep{Ghirardini2019}, and the \textit{Suzaku} measurements in the northern and southern directions \citep{Bautz2009}. In the central region, despite differences in the extraction apertures among the observations, the \textit{EP-FXT} temperature measurements are consistent with previous results within the uncertainties. In the outskirts, the \textit{EP-FXT} temperatures are marginally consistent with the \textit{Chandra} measurements within the uncertainties and are slightly higher than those from \textit{Suzaku} and X-COP, but follow a similar radial trend. Overall, we directly measured the radial temperature profile of A1795 out to $R_{200}$ with full azimuthal coverage and revealed a gradual temperature decline towards the cluster outskirts.

\begin{table*}[t]
\caption{Best-fit spectral parameters of A1795 measured in different annuli.}
\label{tab:a1795_abs1T_annuli}
\centering
{\renewcommand{\arraystretch}{1.20} 
\begin{tabular}{lccccc}
\hline\hline
Annulus (arcmin) & $kT$ (keV) & $Z/Z_{\odot}$ & norm ($10^{-2}$) & C-Statistic (d.o.f.) & C-Statistic/d.o.f. \\
\hline
0.00--0.80 & $4.12^{+0.04}_{-0.04}$& $0.37^{+0.01}_{-0.01}$& $1.610^{+0.008}_{-0.008}$& 6275 (5914)& 1.06\\
0.80--1.50   & $5.10^{+0.06}_{-0.05}$ & $0.27^{+0.02}_{-0.02}$  & $1.482^{+0.007}_{-0.007}$ & 6108 (5892) & 1.04 \\
1.50--2.20   & $5.74^{+0.09}_{-0.08}$ & $0.24^{+0.02}_{-0.02}$  & $1.054^{+0.006}_{-0.006}$ & 5795 (5320) & 1.09\\
2.20--3.20   & $5.99^{+0.09}_{-0.09}$ & $0.16^{+0.02}_{-0.02}$  & $1.088^{+0.006}_{-0.006}$ & 5957 (5318) & 1.12 \\
3.20--4.40   & $6.02^{+0.11}_{-0.10}$ & $0.18^{+0.03}_{-0.03}$  & $0.865^{+0.006}_{-0.006}$ & 5489 (4805) & 1.14 \\
4.40--6.00   & $6.14^{+0.14}_{-0.14}$ & $0.12^{+0.03}_{-0.03}$  & $0.750^{+0.006}_{-0.006}$ & 5325 (4415)& 1.21\\
6.00--8.00   & $5.79^{+0.16}_{-0.18}$ & $0.01^{+0.04}_{-0.01}$  & $0.586^{+0.004}_{-0.006}$ & 5091 (3690)& 1.38\\
-- & $5.88^{+0.15}_{-0.16}$ & 0.10 (fixed) & $0.574^{+0.002}_{-0.002}$ & 5105 (3691)& 1.38\\
8.00--10.00  & $5.06^{+0.21}_{-0.17}$ & $0.00^{+0.82}_{-0.00}$  & $0.367^{+0.002}_{-0.002}$ & 3773 (2651)& 1.42\\
--& $5.30^{+0.21}_{-0.19}$ & 0.10 (fixed) & $0.357^{+0.002}_{-0.002}$ & 3808 (2652)& 1.44\\
10.00--12.00 & $4.29^{+0.26}_{-0.18}$ & $0.00^{+0.01}_{-0.00}$  & $0.214^{+0.002}_{-0.002}$ & 2572 (1787)& 1.44 \\
--& $4.72^{+0.29}_{-0.26}$ & 0.10 (fixed) & $0.206^{+0.002}_{-0.002}$ & 2615 (1788)& 1.46\\
12.00--18.00 & $3.28^{+0.16}_{-0.21}$& $0.00^{+0.00}_{-0.00}$& $0.367^{+0.004}_{-0.003}$& 4632 (3295)& 1.41\\
--& $4.05^{+0.27}_{-0.26}$ & 0.10 (fixed) & $0.344^{+0.004}_{-0.003}$ & 4726 (3296)&1.43\\
18.00--26.00 & $1.73^{+0.30}_{-0.25}$& $0.00^{+0.01}_{-0.00}$& $0.224^{+0.014}_{-0.016}$& 3946 (3601)&1.10\\
--& $2.76^{+0.55}_{-0.37}$ & 0.10 (fixed) & $0.180^{+0.007}_{-0.007}$ & 3974 (3602)& 1.10\\
\hline
\end{tabular}
}
\tablefoot{The spectrum results fitted in the 0.5--10.0 keV range are shown, with the last five annuli showing the results for both fixed and free metal abundance fits. The quoted uncertainties correspond to the 90\% confidence level.}
\end{table*}

\subsection{The brightness residual image}
\begin{figure}[ht!]
   \centering
   \includegraphics[width=\hsize]{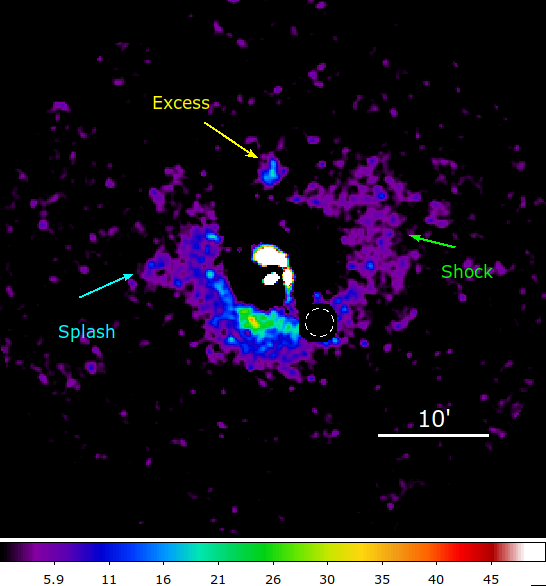}
   \caption{Surface-brightness residual map of A1795 in the 0.3--2.5~keV band. The map covers the same circular region of radius, $R_{200}$, as the yellow circle in Fig.~\ref{fig:2}. The residual values are given in counts. The arrows highlight prominent structures, which are discussed in more detail in Sect.~\ref{sec:discussion}. The white dashed circle indicates the location of the subtracted point source.}
  \label{fig:res_fxt}
\end{figure}

To highlight substructures that are not readily visible in the original source image, we used \texttt{Sherpa} (version 4.17.0)\footnote{\url{https://cxc.cfa.harvard.edu/sherpa/}} to fit a classical double $\beta$ elliptical model to the data. The residual map shown in Fig.~\ref{fig:res_fxt} was obtained by subtracting the best-fit model from the data. From the residual map, we can see a prominent surface-brightness excess spiral structure that originates from the core and extends outward in a clockwise direction towards the south-east. To the east, an additional excess region is present, hereafter referred to as the splash feature \citep{Watson2025}. In addition, a distinct surface-brightness enhancement is present in the north-west and a localised excess is also visible in the northern region. These structures suggest that A1795 has experienced dynamical disturbances and still retains their signatures. The white dashed circle in the south-west of the residual map marks the location of the brightest point source closest to the centre, which was subtracted following the procedure adopted in the \textit{XMM-Newton} analysis by \citet{Bourdin2008}.

\begin{figure*}[t]
\centering

\begin{minipage}{0.49\textwidth}
\centering
\includegraphics[width=\linewidth]{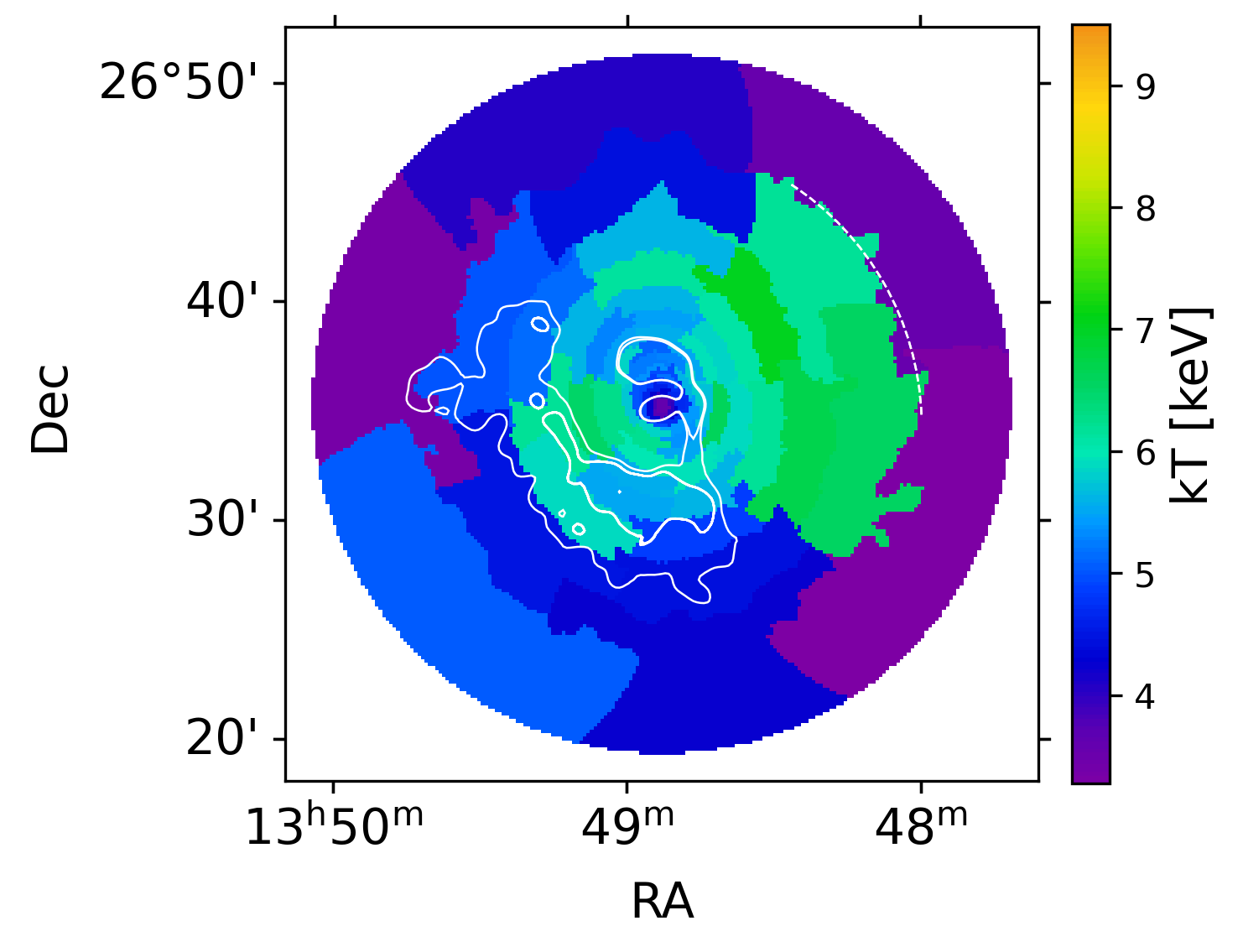}
\end{minipage}\hfill
\begin{minipage}{0.49\textwidth}
\centering
\includegraphics[width=\linewidth]{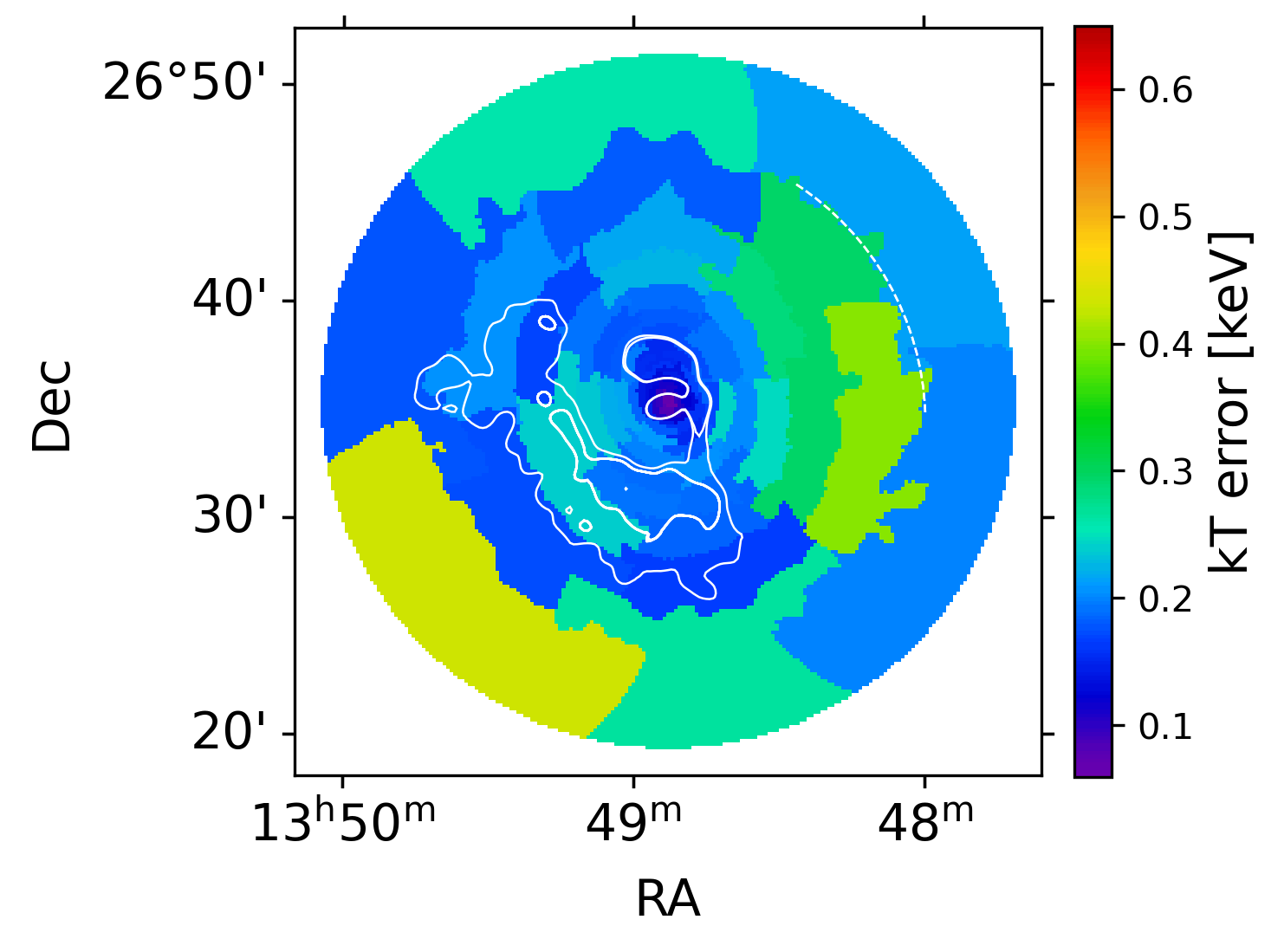}
\end{minipage}

\begin{minipage}{0.49\textwidth}
\centering
\includegraphics[width=\linewidth]{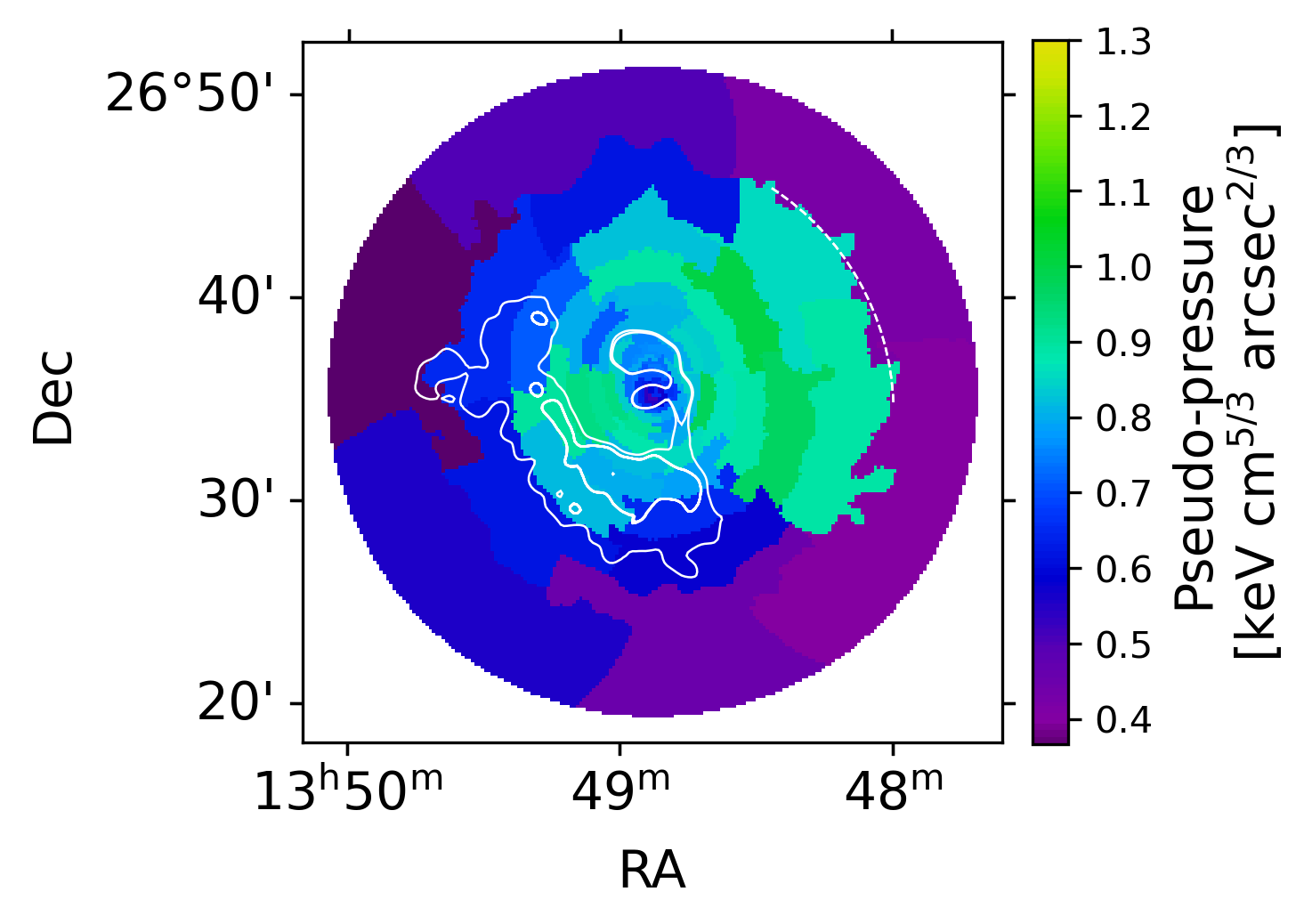}
\end{minipage}\hfill
\begin{minipage}{0.49\textwidth}
\centering
\includegraphics[width=\linewidth]{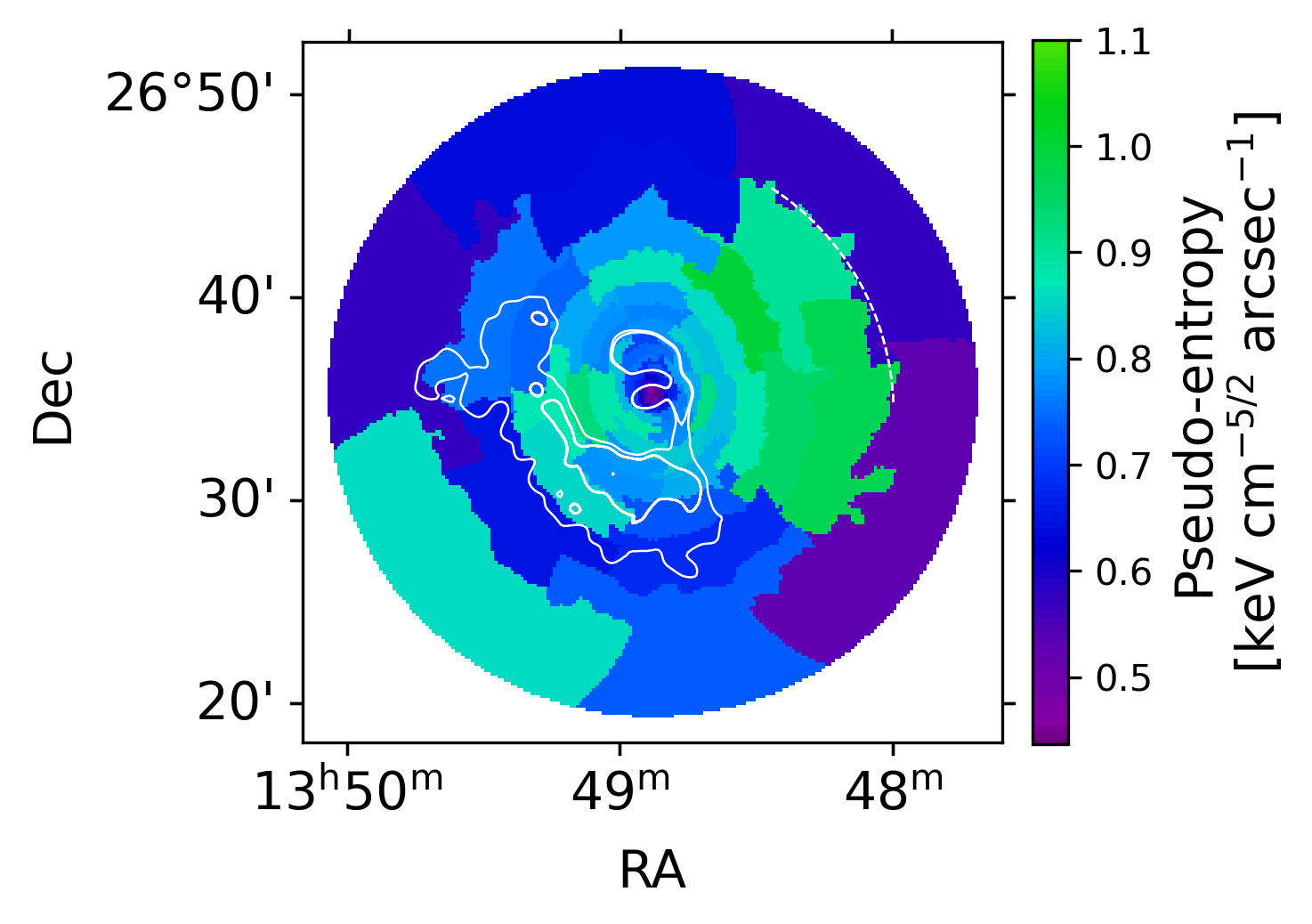}
\end{minipage}

\caption{Thermodynamic maps of A1795 obtained with the contour-binning method. For each region, the spectrum was fitted in the 0.5--10.0~keV band. Top-left: Temperature map. Top-right: Temperature error map. Bottom-left: Pseudo-pressure map. Bottom right: Pseudo-entropy map. The overlaid white contours trace the residual X-ray emission in Fig.~\ref{fig:res_fxt}. The dashed line roughly indicates the location of the high-temperature region in the north-west direction.}
\label{fig:contbin}
\end{figure*}

\subsection{Thermodynamic mapping}\label{subsec:thermodynamic_mapping}
In this work, we conducted thermodynamic mapping to explore the properties of the ICM. In creating these maps, we adopted the contour-binning algorithm \citep{Sanders2006} to divide the field of view into spatial bins with a target S/N of 180, yielding 83 bins in the X-ray image. For each small region, we performed spectral fitting using the \texttt{TBABS*APEC} model in the 0.5--10.0~keV energy range, adopting the same parameter settings as in Sect.~\ref{subsec:radial_temperature_profile}, with the abundance allowed to vary freely. Based on the best-fit temperature parameters and normalisation values, we calculated the pseudo-pressure and pseudo-entropy maps using the relations \( P = kT n_e \) and \( S = kT n_e^{-2/3} \), respectively \citep{sasaki2016}. Here, \( n_e \) denotes the electron density and \( kT \) is the temperature. Since the normalisation derived from spectral fitting is proportional to the square of the electron density, we express the pressure and entropy as \(P \propto kT\,\mathrm{norm}^{0.5}\) and \(S \propto kT\,\mathrm{norm}^{-1/3}\) \citep{Fabian2006}, as shown in Fig.~\ref{fig:contbin}.

The temperature map shows a pronounced high-temperature region in the north-west, corresponding to the excess surface-brightness feature seen in the residual image. A low-temperature band extending clockwise from the centre coincides with the spiral-like excess structure in the residual map; at a given radius, the temperatures along this direction are systematically lower than those at the same radius in other azimuths. The \textit{Chandra} temperature map presented by \citet{Ehlert2015} is consistent with our \textit{EP-FXT} measurements in the central region: A1795 exhibits a cool-core temperature structure, with low temperatures in the core and temperatures increasing with radius. They also detected a temperature drop in the south-eastern region associated with part of the spiral structure that we detect, but their limited FOV did not allow them to trace the whole structure. With the larger FOV of \textit{EP-FXT}, we were able to reveal the thermal structure over a larger area and its spatial pattern broadly follows the spiral morphology highlighted in the surface-brightness residual map.

In the pseudo-pressure and pseudo-entropy maps, the hottest region in the north-west also exhibits the highest pressure and entropy, suggesting the possible presence of a shock. In contrast, the low-temperature region associated with the spiral structure shows a reduced entropy distribution. These low-temperature and low-entropy signatures further support an interpretation in terms of sloshing cold fronts, as previously discussed for the core of A1795 \citep{Markevitch2002,Ehlert2015}.

\begin{figure}[ht!]
   \centering
   \includegraphics[width=\hsize]{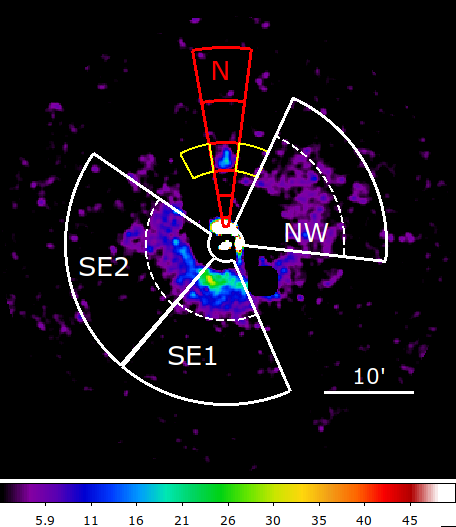}
   \caption{Surface-brightness residual map of A1795 in the 0.3--2.5~keV band. The four regions of interest are marked. The white dashed lines within the sector regions indicate the locations of the surface-brightness discontinuities.}
  \label{fig:res_fxt_region}
\end{figure}

\begin{figure}[ht!]
   \centering
   \includegraphics[width=\hsize]{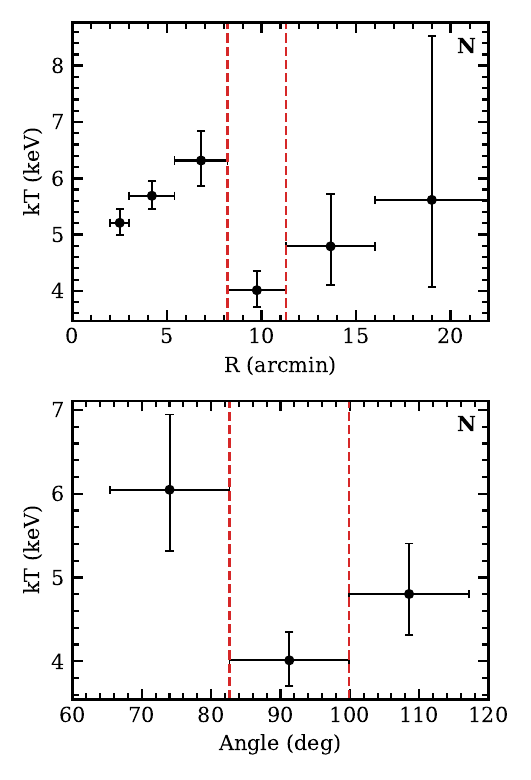}
   \caption{Temperature profiles extracted from the northern sectors shown in Fig.~\ref{fig:res_fxt_region}. The upper and lower panels show the temperature versus radius and versus azimuthal angle (from west to east) within the same radial interval, respectively. The radial interval marked by the red dashed lines corresponds to the inner and outer boundaries of the surface-brightness excess region in the northern residual map. Uncertainties are quoted at the 90\% confidence level.}
  \label{fig:n_excess}
\end{figure}

\section{Discussion}\label{sec:discussion}
\subsection{Northern excess}
At $\sim 8\arcmin$ north of the cluster centre, we identified a prominent region of excess X-ray emission in the surface-brightness residual map. To quantify this feature, as illustrated in Fig.~\ref{fig:res_fxt_region}, we defined six fan-shaped sectors extending northward from the cluster centre (red regions) and (over the same radial range) selected three comparison sectors distributed from west to east (yellow regions). As shown in Fig.~\ref{fig:n_excess}, the sector associated with the enhanced residual emission exhibits systematically lower temperatures than the other sectors within the radial interval marked by the red dashed lines, in both sets of temperature profiles.

We further cross-matched the sky position of this excess region with the optical cluster catalogues compiled by \citet{WH2015,WH2024}. The excess is spatially coincident with the galaxy cluster J134855.9+264410 (J2000: $\alpha=13^{\rm h}48^{\rm m}55.896^{\rm s}$, $\delta=+26^\circ44^\prime10.23^{\prime\prime}$), which lies at a spectroscopic redshift of $z=0.248$, and is characterised by $M_{500}=1.21\times10^{14} M_{\odot}$ and $R_{500}=0.75~\mathrm{Mpc}$ \citep{WH2024}. Given the significantly lower best-fit temperature of the excess compared to the surrounding ICM of A1795 and its positional association with J134855.9+264410, we favour the interpretation that the excess is associated with a background galaxy cluster. 

At $z=0.248$, the $R_{500}$ of the background cluster corresponds to an angular radius of approximately $3.2\arcmin$. We extracted its spectrum from a circular region of radius $R_{500}$ and used nearby source-free regions of equal area for background subtraction, yielding a best-fit temperature of $kT = 3.27^{+0.87}_{-0.68}$ keV. While this cluster partially overlaps with A1795, its impact is localised, as evidenced by the surface-brightness enhancement in Fig.~\ref{fig:2} and the corresponding temperature drop in Fig.~\ref{fig:contbin}. Furthermore, we verified that excluding this region does not significantly alter the global ICM properties of A1795.

\subsection{The shock front and cold front}
\begin{figure}[ht!]
   \centering
   \includegraphics[width=\hsize]{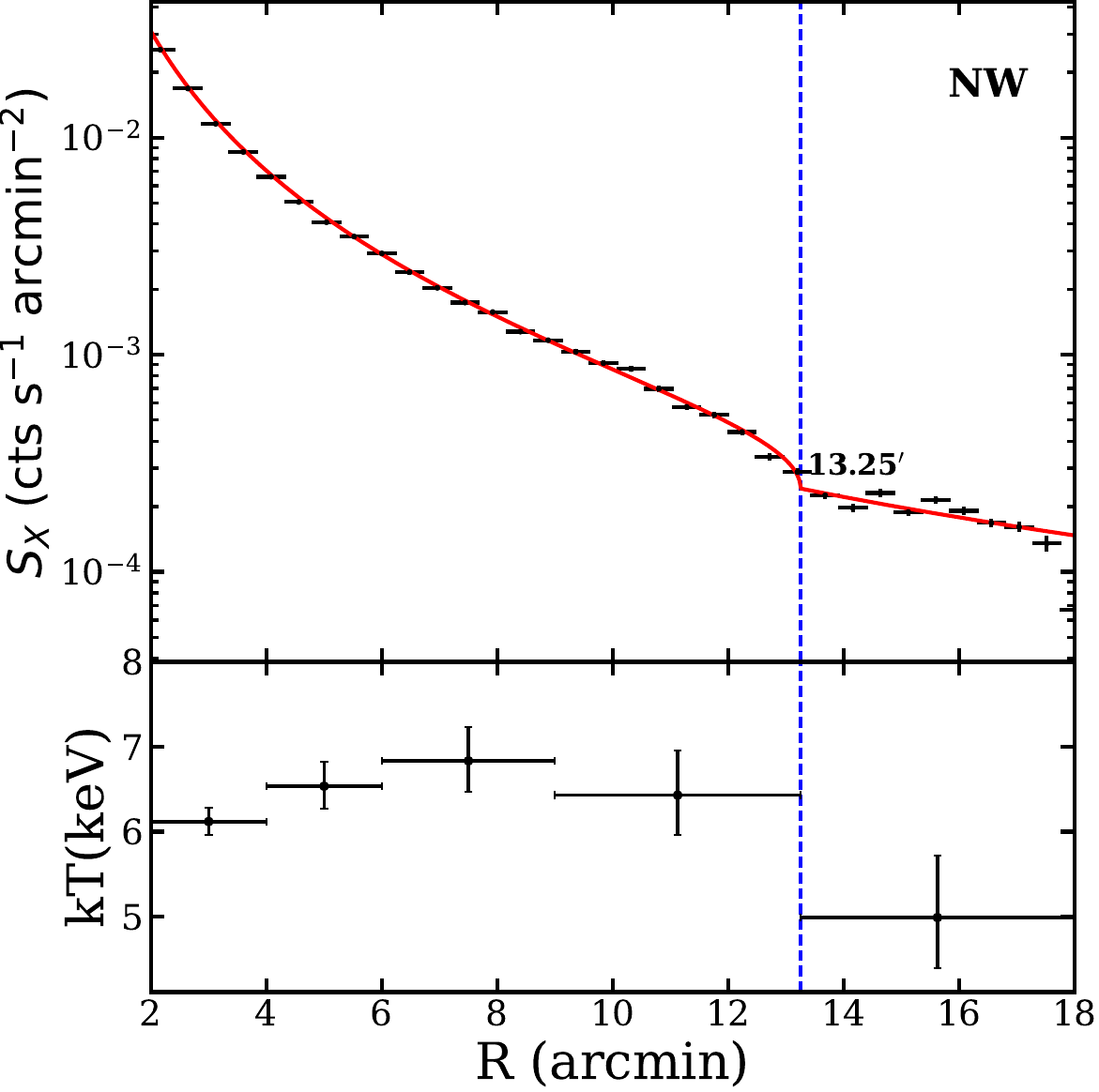}
   \caption{Surface-brightness and temperature profiles shown in the upper and lower panels, respectively, for the NW sector. The red curve indicates the best-fit broken power-law model, and the blue dashed line marks the radius of the discontinuity.}
  \label{fig:nw}
\end{figure}

\begin{figure}[ht!]
   \centering
   \includegraphics[width=\hsize]{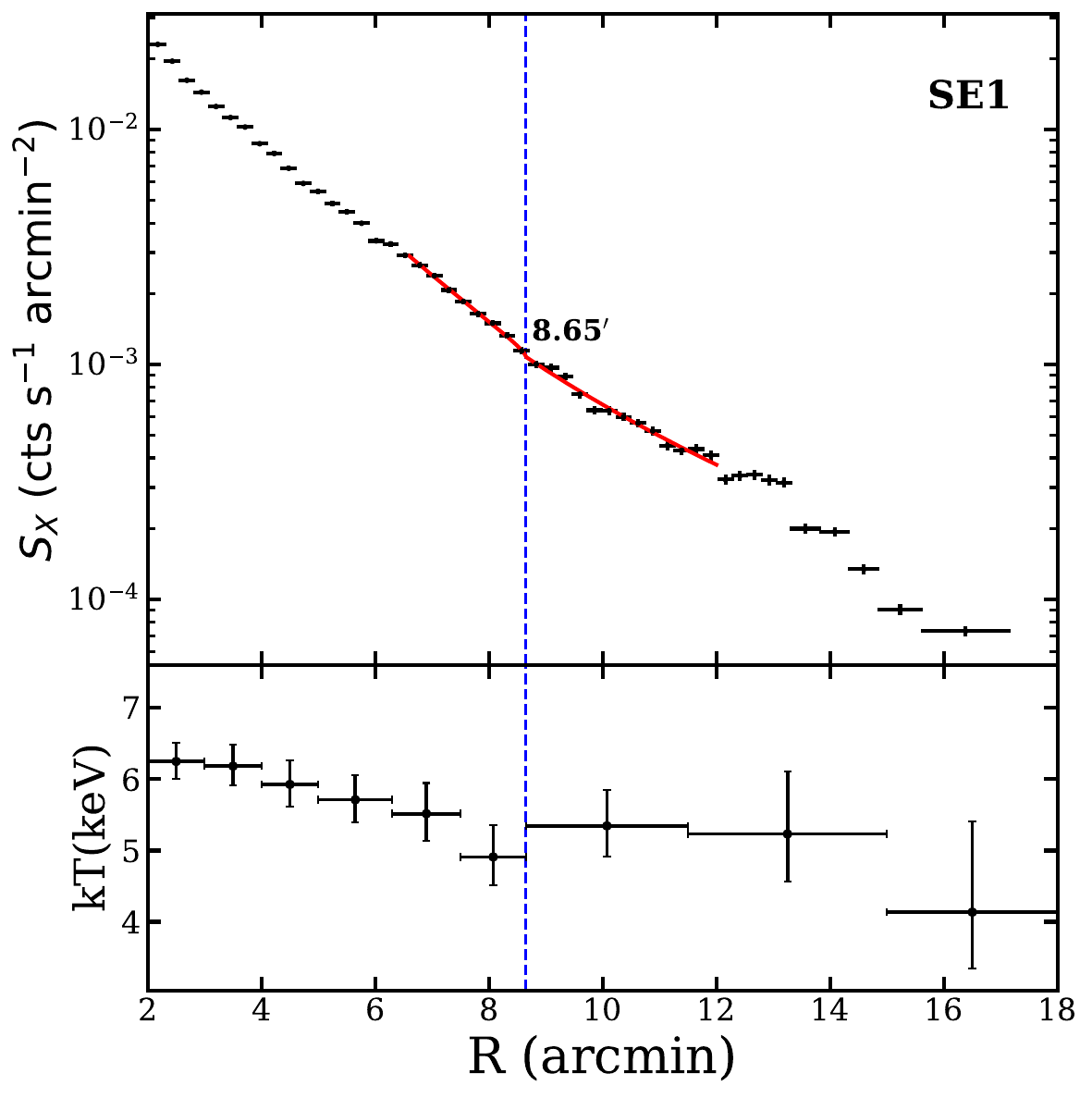}
   \includegraphics[width=\hsize]{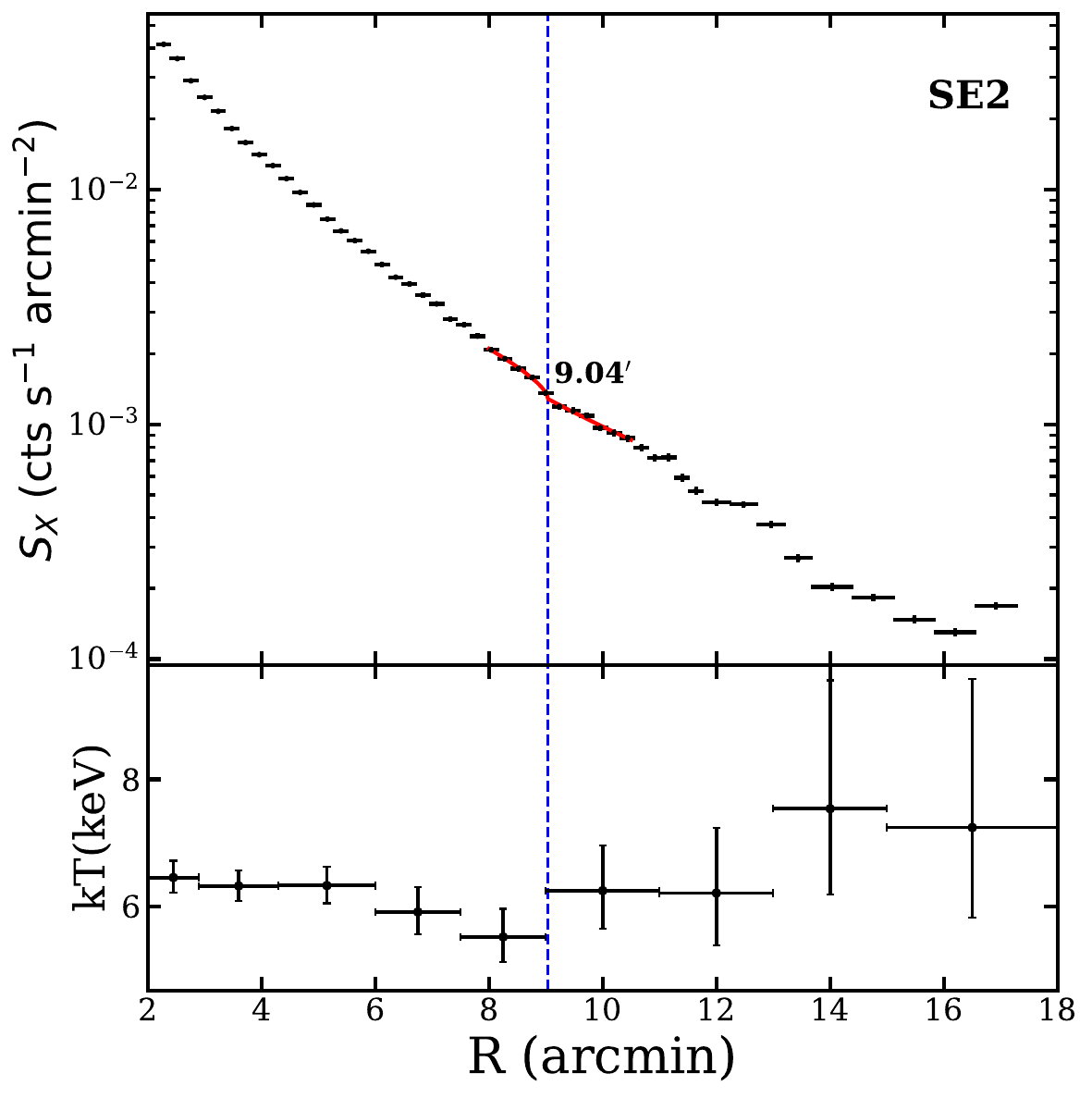}
   \caption{Same as Fig.~\ref{fig:nw}, but for the SE1 (upper) and SE2 (bottom) sectors.}
  \label{fig:SE}
\end{figure}

Although the X-ray morphology of A1795 appears highly symmetric and globally regular, the residual and thermodynamic maps reveal significant substructures. Therefore, we selected three sectors that cover the spiral feature and the north-west brightness excess for detailed radial surface-brightness and spectral analyses. The sector definitions are shown in Fig.~\ref{fig:res_fxt_region}. For each sector, we extract radial surface-brightness profiles and fit them with a projected broken power-law model. We also extracted spectra and fit them with the \texttt{TBABS*APEC} model to derive the corresponding radial temperature profiles (Figs.~\ref{fig:nw} and \ref{fig:SE}).

In the north-western (NW) direction, we detect a surface-brightness edge at $r_{\rm edge}=13.25\pm0.03$ arcmin (Fig.~\ref{fig:nw}). To test whether this structure could be associated with shock heating, we estimated its Mach number. One approach is based on the density compression factor, $C$, \citep{Mirakhor2023} via
\begin{equation}
M_c=\left[\frac{2C}{(\gamma+1)-(\gamma-1)C}\right]^{1/2},
\label{eq:mc}
\end{equation}
where $C \equiv \rho_2/\rho_1$ is the ratio of the post-shock density $\rho_2$ to the pre-shock density $\rho_1$, and $\gamma$ is the adiabatic index, taken to be $5/3$ \citep{sasaki2016}. For the NW edge, we obtained $C = 1.673\pm0.073$, corresponding to $M_c = 1.47^{+0.06}_{-0.05}$, suggesting that this edge may be consistent with a weak shock.

We can also estimate the Mach number from the temperature jump. For an ideal monatomic gas ($\gamma=5/3$), the Mach number can be expressed in terms of the temperature ratio, $T_2/T_1$, as \citep{Sanders2016MNRAS}:
\begin{equation}
M_T =
\left[
\frac{(8\,T_2/T_1-7)+\sqrt{(8\,T_2/T_1-7)^2+15}}{5}
\right]^{1/2},
\label{eq:mt}
\end{equation}
where $T_1$ and $T_2$ are the pre-shock and post-shock temperatures, respectively. Using the two radial temperature bins immediately inside and outside $r_{\rm edge}$, we obtain $T_{2}=6.43^{+0.53}_{-0.47}\,\mathrm{keV}$ and $T_{1}=4.99^{+0.73}_{-0.59}\,\mathrm{keV}$, yielding $T_{2}/T_{1}=1.29$ and $M_{T}=1.30^{+0.29}_{-0.25}$. The temperature jump provides additional evidence for a shock front. These two independent calculations of the Mach number are consistent within the $1\sigma$ uncertainties.

In the two south-eastern sectors associated with the spiral structure (SE1 and SE2), we identified weak edge-like signatures. In SE1, a mild surface-brightness discontinuity is seen at $R \approx 8.65\arcmin$, where the temperature increases from $kT=4.9^{+0.4}_{-0.4}\,\mathrm{keV}$ to $kT=5.3^{+0.5}_{-0.4}\,\mathrm{keV}$. In SE2, a similar feature appears at $R \approx 9.04\arcmin$, with the temperature increasing from $kT=5.5^{+0.4}_{-0.4}\,\mathrm{keV}$ to $kT=6.2^{+0.7}_{-0.6}\,\mathrm{keV}$. Although the changes are not significant, the residual map shows a transition from excess to deficit across the same locations, while the thermodynamic maps reveal low-temperature, low-entropy gas along the spiral pattern. These characteristics are consistent with the typical observational signatures of cold fronts. Overall, we interpret the spiral structure as cold fronts induced by gas sloshing.

\begin{figure}[ht!]
   \centering
   \includegraphics[width=\hsize]{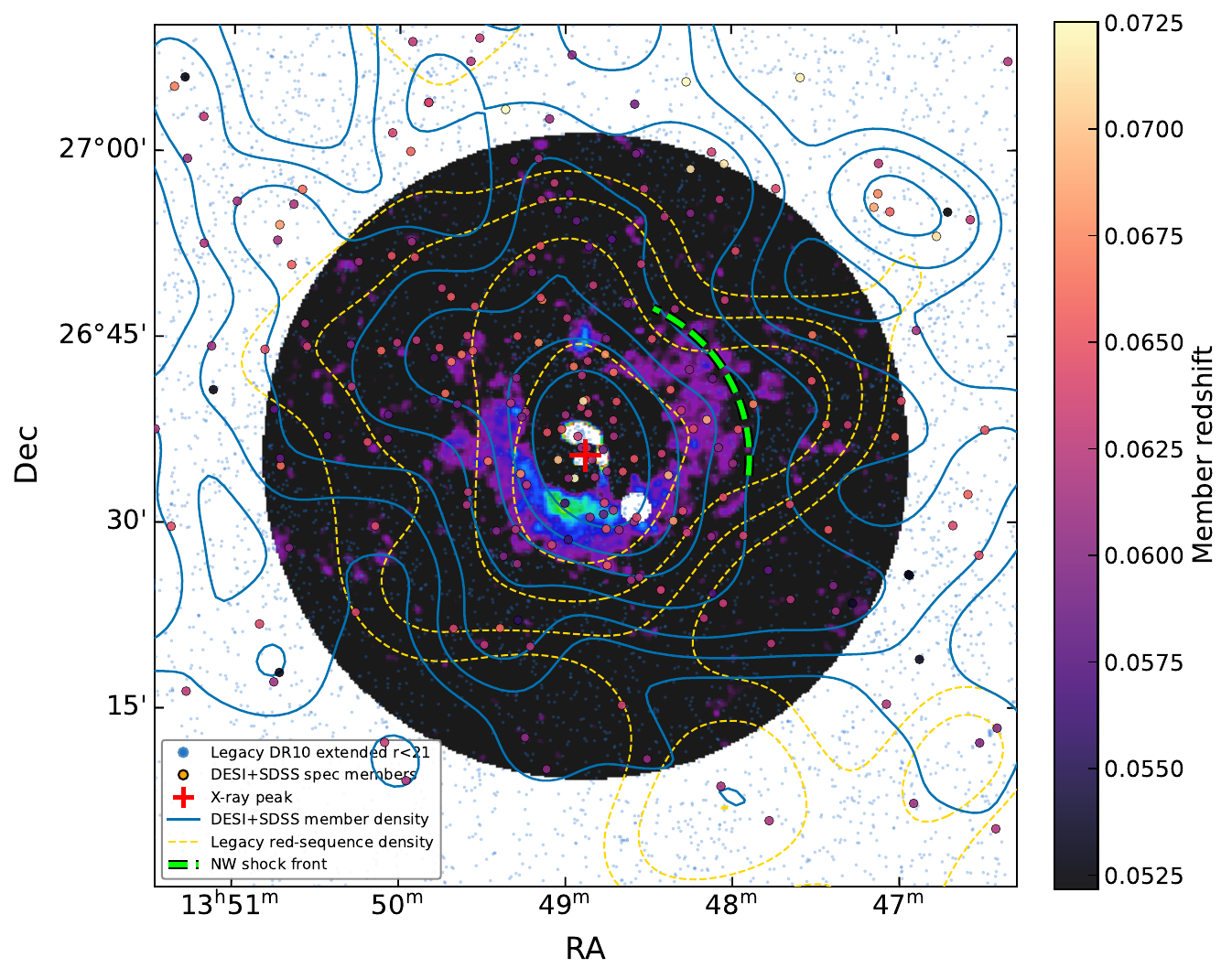}
   \caption{
   Comparison between the X-ray substructures and the optical galaxy distribution in A1795. The X-ray residual map is overlaid with Legacy DR10 extended sources with \(r<21\) (small blue points) and DESI+SDSS candidate spectroscopic members (filled circles colour-coded by redshift). Blue solid contours show the smoothed projected number density of the DESI+SDSS candidate spectroscopic members, while yellow dashed contours show that of the Legacy red-sequence proxy sample.}
  \label{fig:optical_ds_shock}
\end{figure}

\subsection{A binary-merger scenario for substructures}
Combining the X-ray surface-brightness residual map with the 2D thermodynamic maps and supported by the radial sector analysis, we found that the sloshing spiral arm extending from the core towards the south-east traces low-temperature, low-entropy gas. In the surface-brightness enhanced region to the north-west, we identified an arc-like feature consistent with a shock front, marked by elevated temperatures.

The overall observational substructures in A1795, including the sloshing spiral structure, the splash-like feature, and the shock front, are broadly consistent with the binary-merger scenario proposed for A2029 \citep{ZuHone2011,ZuHone2018,Watson2025}. In this scenario, the perturbing subcluster initially passes north of the primary cluster core and moves eastward. The associated gravitational disturbance induces gas sloshing in the central potential well dominated by dark matter, producing a spiral-like cold-front pattern. Given that the subcluster is on a bound orbit, it subsequently approaches again from the east and moves westward. The second core passage may produce a splash feature to the south-east, similar to the wake of material observed in A2029 that was left behind by a perturbed subcluster \citep{Watson2025, ZuHone2018}. During its return passage through the ICM, the perturber's high velocity relative to the ambient gas can compress and dissipatively heat the gas, thereby driving a shock front. 

Based on the merger scenario discussed above, the perturbing system associated with the putative shock front might be expected to lie towards the north-west of the cluster centre, where it could have driven the shock front \citep{ZuHone2011,ZuHone2018,Watson2025}. We searched for a present-day X-ray or optical counterpart in this region. In the X-ray data, we do not identify a clear present-day subhalo at the shock location. In the optical, we compared the X-ray substructures with the projected galaxy distribution. We selected a broad sample of 360 candidate spectroscopic members from the combined DESI DR1 and SDSS DR17 catalogues within \(0.052 < z < 0.073\) \citep{DESI2026,Abdurro2022}. As complementary photometric tracers, we used 17,067 Legacy DR10 extended-source candidates with \(r<21\) \citep{Dey2019}, from which we further defined a red-sequence proxy subsample of 2,502 sources. From these optical samples, we constructed smoothed projected galaxy number-density contours, as shown in Fig.~\ref{fig:optical_ds_shock}. Although the individual candidate spectroscopic members do not show a clear galaxy group near the putative shock front, the number-density contours are slightly extended towards the shock region. Taken together, these results suggest that even apparently relaxed galaxy clusters such as A1795 may still retain residual signatures of past dynamical activity.

\section{Conclusions}\label{sec:conclusion}

Based on deep \textit{EP-FXT} observations with a total exposure of 480~ks, we performed a systematic study of the X-ray imaging and spectral properties of the galaxy cluster A1795. By combining 2D thermodynamic maps with radial-profile measurements in azimuthal sectors, we further investigated its internal substructures. Our main conclusions are summarised as follows:

1. Benefiting from the large field of view and low instrumental background of \textit{EP-FXT}, we obtained high-S/N X-ray images of A1795, with coverage extending out to $R_{200}$. The cluster exhibits a globally regular X-ray morphology, with a surface-brightness distribution characteristic of cool-core clusters.

2. \textit{EP-FXT} directly measures the radial temperature profile of A1795 out to $R_{200}$ with full azimuthal coverage. The profile increases with radius within $6\arcmin$, reaches a peak, and then gradually declines towards larger radii, providing additional constraints on the thermal properties in the cluster outskirts.

3. The surface-brightness residuals, thermodynamic maps, and radial sector profiles consistently show that the spiral arm extending from the core towards the south-east is associated with low-temperature, low-entropy gas, and is consistent with sloshing-induced cold fronts. The brightness-enhanced region to the north-west exhibits an arc-like high-temperature feature consistent with a shock front.

Based on the detected substructures, we suggest that they can be interpreted within a binary merger scenario. In particular, the initial passage of the perturbing subcluster triggers core gas sloshing, while its subsequent return may drive a shock front towards the north-west in the ICM.

\begin{acknowledgements}
This work is based on the data obtained with Einstein Probe, a space mission supported by the Strategic Priority Program on Space Science of Chinese Academy of Sciences, in collaboration with the European Space Agency, the Max-Planck-Institute for extraterrestrial Physics (Germany), and the Centre National d'Etudes Spatiales (France). We acknowledge support from the National Key R\&D Program of China No.~2025YFF0511104 and the International Partnership Program of Chinese Academy of Sciences (Grant No.~013GJHZ2024015FN). Y.H. acknowledges the support from the National Natural Science Foundation of China No.~12573003. SA acknowledges PRIN-MIUR grant 20228B938N ``Mass and selection biases of galaxy clusters: a multi-probe approach'' funded by the European Union Next Generation EU, Mission 4 Component 1, CUP C53D2300092 0006.

\end{acknowledgements}

\bibliography{ref}{}

@ARTICLE{Tamura2001,
       author = {{Tamura}, T. and {Kaastra}, J.~S. and {Peterson}, J.~R. and {Paerels}, F.~B.~S. and {Mittaz}, J.~P.~D. and {Trudolyubov}, S.~P. and {Stewart}, G. and {Fabian}, A.~C. and {Mushotzky}, R.~F. and {Lumb}, D.~H. and {Ikebe}, Y.},
        title = "{X-ray spectroscopy of the cluster of galaxies Abell 1795 with XMM-Newton}",
      journal = {\aap},
         year = 2001,
        month = jan,
       volume = {365},
        pages = {L87-L92},
          doi = {10.1051/0004-6361:20000038},
archivePrefix = {arXiv},
       eprint = {astro-ph/0010362},
 primaryClass = {astro-ph},
       adsurl = {https://ui.adsabs.harvard.edu/abs/2001A&A...365L..87T}
}

@article{AndersGrevesse1989,
title = {Abundances of the elements: Meteoritic and solar},
journal = {Geochimica et Cosmochimica Acta},
volume = {53},
number = {1},
pages = {197-214},
year = {1989},
issn = {0016-7037},
doi = {https://doi.org/10.1016/0016-7037(89)90286-X},
url = {https://www.sciencedirect.com/science/article/pii/001670378990286X},
author = {Edward Anders and Nicolas Grevesse}
}

@ARTICLE{Zheng2025,
       author = {{Zheng}, X. and {Jia}, S. and {Li}, C. and {Chen}, Y. and {Yu}, H. and {Feng}, H. and {Xu}, D. and {Liu}, A. and {Song}, L. and {Liu}, C. and {Lu}, F. and {Zhang}, S. and {Yuan}, W. and {Sanders}, J. and {Wang}, J. and {Chen}, T. and {Cui}, C. and {Cui}, W. and {Feng}, W. and {Gao}, N. and {Guan}, J. and {Han}, D. and {Hou}, D. and {Hu}, H. and {Huang}, M. and {Huo}, J. and {Jin}, C. and {Li}, M. and {Li}, W. and {Liu}, Y. and {Luo}, L. and {Ma}, J. and {Ou}, G. and {Pan}, H. and {Wang}, H. and {Wang}, Ji. and {Wang}, Ju. and {Wang}, Y. and {Xu}, J. and {Xu}, Y. and {Yang}, X. and {Yang}, Y. and {Zhang}, H. and {Zhang}, J. and {Zhang}, M. and {Zhang}, Z. and {Zhao}, H. and {Zhao}, X. and {Zhao}, Z. and {Zhu}, P. and {Zhu}, Y.},
        title = "{Imaging-spectroscopy diagnosis of the giant sloshing spiral in the Virgo cluster with the Einstein Probe Follow-up X-ray Telescope}",
      journal = {\aap},
         year = 2025,
        month = aug,
       volume = {700},
          eid = {A248},
        pages = {A248},
          doi = {10.1051/0004-6361/202554719},
archivePrefix = {arXiv},
       eprint = {2507.07412},
 primaryClass = {astro-ph.HE},
       adsurl = {https://ui.adsabs.harvard.edu/abs/2025A&A...700A.248Z}
}

@ARTICLE{Ebeling1998,
       author = {{Ebeling}, H. and {Edge}, A.~C. and {Bohringer}, H. and {Allen}, S.~W. and {Crawford}, C.~S. and {Fabian}, A.~C. and {Voges}, W. and {Huchra}, J.~P.},
        title = "{The ROSAT Brightest Cluster Sample - I. The compilation of the sample and the cluster log N-log S distribution}",
      journal = {\mnras},
         year = 1998,
        month = dec,
       volume = {301},
       number = {4},
        pages = {881-914},
          doi = {10.1046/j.1365-8711.1998.01949.x},
archivePrefix = {arXiv},
       eprint = {astro-ph/9812394},
 primaryClass = {astro-ph},
       adsurl = {https://ui.adsabs.harvard.edu/abs/1998MNRAS.301..881E}
}

@ARTICLE{Markevitch2001,
       author = {{Markevitch}, M. and {Vikhlinin}, A. and {Mazzotta}, P.},
        title = "{Nonhydrostatic Gas in the Core of the Relaxed Galaxy Cluster A1795}",
      journal = {\apjl},
         year = 2001,
        month = dec,
       volume = {562},
       number = {2},
        pages = {L153-L156},
          doi = {10.1086/337973},
archivePrefix = {arXiv},
       eprint = {astro-ph/0108520},
 primaryClass = {astro-ph},
       adsurl = {https://ui.adsabs.harvard.edu/abs/2001ApJ...562L.153M}
}

@ARTICLE{Fabian2001,
       author = {{Fabian}, A.~C. and {Sanders}, J.~S. and {Ettori}, S. and {Taylor}, G.~B. and {Allen}, S.~W. and {Crawford}, C.~S. and {Iwasawa}, K. and {Johnstone}, R.~M.},
        title = "{Chandra imaging of the X-ray core of Abell 1795}",
      journal = {\mnras},
         year = 2001,
        month = feb,
       volume = {321},
       number = {2},
        pages = {L33-L36},
          doi = {10.1046/j.1365-8711.2001.04243.x},
archivePrefix = {arXiv},
       eprint = {astro-ph/0011547},
 primaryClass = {astro-ph},
       adsurl = {https://ui.adsabs.harvard.edu/abs/2001MNRAS.321L..33F}
}

@ARTICLE{Kovacs2023,
       author = {{Kov{\'a}cs}, Orsolya Eszter and {Zhu}, Zhenlin and {Werner}, Norbert and {Simionescu}, Aurora and {Bogd{\'a}n}, {\'A}kos},
        title = "{Outskirts of Abell 1795: Probing gas clumping in the intracluster medium}",
      journal = {\aap},
         year = 2023,
        month = oct,
       volume = {678},
          eid = {A91},
        pages = {A91},
          doi = {10.1051/0004-6361/202347201},
archivePrefix = {arXiv},
       eprint = {2306.10101},
 primaryClass = {astro-ph.GA},
       adsurl = {https://ui.adsabs.harvard.edu/abs/2023A&A...678A..91K}
}

@ARTICLE{Bautz2009,
       author = {{Bautz}, Marshall W. and {Miller}, Eric D. and {Sanders}, Jeremy S. and {Arnaud}, Keith A. and {Mushotzky}, Richard F. and {Porter}, F. Scott and {Hayashida}, Kiyoshi and {Henry}, J. Patrick and {Hughes}, John P. and {Kawaharada}, Madoka and {Makashima}, Kazuo and {Sato}, Mitsuhiro and {Tamura}, Takayuki},
        title = "{Suzaku Observations of Abell 1795: Cluster Emission to r$_{200}$}",
      journal = {\pasj},
         year = 2009,
        month = oct,
       volume = {61},
        pages = {1117},
          doi = {10.1093/pasj/61.5.1117},
archivePrefix = {arXiv},
       eprint = {0906.3515},
 primaryClass = {astro-ph.CO},
       adsurl = {https://ui.adsabs.harvard.edu/abs/2009PASJ...61.1117B}
}

@ARTICLE{Ehlert2015,
       author = {{Ehlert}, Steven and {McDonald}, Michael and {David}, Laurence P. and {Miller}, Eric D. and {Bautz}, Mark W.},
        title = "{A Very Deep Chandra Observation of A1795: The Cold Front and Cooling Wake}",
      journal = {\apj},
         year = 2015,
        month = feb,
       volume = {799},
       number = {2},
          eid = {174},
        pages = {174},
          doi = {10.1088/0004-637X/799/2/174},
archivePrefix = {arXiv},
       eprint = {1406.4352},
 primaryClass = {astro-ph.GA},
       adsurl = {https://ui.adsabs.harvard.edu/abs/2015ApJ...799..174E}
}

@ARTICLE{Ettori2002,
       author = {{Ettori}, S. and {Fabian}, A.~C. and {Allen}, S.~W. and {Johnstone}, R.~M.},
        title = "{Deep inside the core of Abell 1795: the Chandra view}",
      journal = {\mnras},
         year = 2002,
        month = apr,
       volume = {331},
       number = {3},
        pages = {635-648},
          doi = {10.1046/j.1365-8711.2002.05212.x},
archivePrefix = {arXiv},
       eprint = {astro-ph/0111586},
 primaryClass = {astro-ph},
       adsurl = {https://ui.adsabs.harvard.edu/abs/2002MNRAS.331..635E}
}

@INPROCEEDINGS{Gu2012,
       author = {{Gu}, Liyi and {Xu}, Haiguang and {Makishima}, Kazuo},
        title = "{Two-phase ICM in the central region of the rich cluster of galaxies Abell 1795: A joint Chandra, XMM-Newton, and Suzaku view}",
    booktitle = {Suzaku 2011: Exploring the X-ray Universe: Suzaku and Beyond},
         year = 2012,
       editor = {{Petre}, Rob and {Mitsuda}, Kazuhisa and {Angelini}, Lorella},
       series = {American Institute of Physics Conference Series},
       volume = {1427},
        month = mar,
    publisher = {AIP},
        pages = {334-335},
          doi = {10.1063/1.3696230},
       adsurl = {https://ui.adsabs.harvard.edu/abs/2012AIPC.1427..334G}
}

@ARTICLE{Xu1998,
       author = {{Xu}, Haiguang and {Makishima}, Kazuo and {Fukazawa}, Yasushi and {Ikebe}, Yasushi and {Kikuchi}, Ken'ichi and {Ohashi}, Takaya and {Tamura}, Takayuki},
        title = "{Discovery of the Central Excess Brightness in Hard X-Rays in the Cluster of Galaxies Abell 1795}",
      journal = {\apj},
         year = 1998,
        month = jun,
       volume = {500},
       number = {2},
        pages = {738-749},
          doi = {10.1086/305744},
archivePrefix = {arXiv},
       eprint = {astro-ph/9806279},
 primaryClass = {astro-ph},
       adsurl = {https://ui.adsabs.harvard.edu/abs/1998ApJ...500..738X}
}

@ARTICLE{Buote1996,
       author = {{Buote}, David A. and {Tsai}, John C.},
        title = "{Quantifying the Morphologies and Dynamical Evolution of Galaxy Clusters. II. Application to a Sample of ROSAT Clusters}",
      journal = {\apj},
         year = 1996,
        month = feb,
       volume = {458},
        pages = {27},
          doi = {10.1086/176790},
archivePrefix = {arXiv},
       eprint = {astro-ph/9504046},
 primaryClass = {astro-ph},
       adsurl = {https://ui.adsabs.harvard.edu/abs/1996ApJ...458...27B}
}

@ARTICLE{Kokotanekov2018,
       author = {{Kokotanekov}, G. and {Wise}, M.~W. and {de Vries}, M. and {Intema}, H.~T.},
        title = "{Signatures of multiple episodes of AGN activity in the core of Abell 1795}",
      journal = {\aap},
         year = 2018,
        month = oct,
       volume = {618},
          eid = {A152},
        pages = {A152},
          doi = {10.1051/0004-6361/201833222},
archivePrefix = {arXiv},
       eprint = {1807.11520},
 primaryClass = {astro-ph.GA},
       adsurl = {https://ui.adsabs.harvard.edu/abs/2018A&A...618A.152K}
}

@ARTICLE{Vikhlinin2006,
       author = {{Vikhlinin}, A. and {Kravtsov}, A. and {Forman}, W. and {Jones}, C. and {Markevitch}, M. and {Murray}, S.~S. and {Van Speybroeck}, L.},
        title = "{Chandra Sample of Nearby Relaxed Galaxy Clusters: Mass, Gas Fraction, and Mass-Temperature Relation}",
      journal = {\apj},
         year = 2006,
        month = apr,
       volume = {640},
       number = {2},
        pages = {691-709},
          doi = {10.1086/500288},
archivePrefix = {arXiv},
       eprint = {astro-ph/0507092},
 primaryClass = {astro-ph},
       adsurl = {https://ui.adsabs.harvard.edu/abs/2006ApJ...640..691V}
}

@ARTICLE{Vikhlinin2014,
       author = {{Vikhlinin}, A.~A. and {Kravtsov}, A.~V. and {Markevich}, M.~L. and {Sunyaev}, R.~A. and {Churazov}, E.~M.},
        title = "{Clusters of galaxies}",
      journal = {Physics Uspekhi},
         year = 2014,
        month = apr,
       volume = {57},
       number = {4},
          eid = {317-341},
        pages = {317-341},
          doi = {10.3367/UFNe.0184.201404a.0339},
       adsurl = {https://ui.adsabs.harvard.edu/abs/2014PhyU...57..317V}
}

@ARTICLE{Markevitch2007,
       author = {{Markevitch}, Maxim and {Vikhlinin}, Alexey},
        title = "{Shocks and cold fronts in galaxy clusters}",
      journal = {\physrep},
         year = 2007,
        month = may,
       volume = {443},
       number = {1},
        pages = {1-53},
          doi = {10.1016/j.physrep.2007.01.001},
archivePrefix = {arXiv},
       eprint = {astro-ph/0701821},
 primaryClass = {astro-ph},
       adsurl = {https://ui.adsabs.harvard.edu/abs/2007PhR...443....1M}
}

@ARTICLE{Walker2017,
       author = {{Walker}, S.~A. and {Hlavacek-Larrondo}, J. and {Gendron-Marsolais}, M. and {Fabian}, A.~C. and {Intema}, H. and {Sanders}, J.~S. and {Bamford}, J.~T. and {van Weeren}, R.},
        title = "{Is there a giant Kelvin-Helmholtz instability in the sloshing cold front of the Perseus cluster?}",
      journal = {\mnras},
         year = 2017,
        month = jun,
       volume = {468},
       number = {2},
        pages = {2506-2516},
          doi = {10.1093/mnras/stx640},
archivePrefix = {arXiv},
       eprint = {1705.00011},
 primaryClass = {astro-ph.CO},
       adsurl = {https://ui.adsabs.harvard.edu/abs/2017MNRAS.468.2506W}
}

@ARTICLE{Eckert2019,
       author = {{Eckert}, D. and {Ghirardini}, V. and {Ettori}, S. and {Rasia}, E. and {Biffi}, V. and {Pointecouteau}, E. and {Rossetti}, M. and {Molendi}, S. and {Vazza}, F. and {Gastaldello}, F. and {Gaspari}, M. and {De Grandi}, S. and {Ghizzardi}, S. and {Bourdin}, H. and {Tchernin}, C. and {Roncarelli}, M.},
        title = "{Non-thermal pressure support in X-COP galaxy clusters}",
      journal = {\aap},
         year = 2019,
        month = jan,
       volume = {621},
          eid = {A40},
        pages = {A40},
          doi = {10.1051/0004-6361/201833324},
archivePrefix = {arXiv},
       eprint = {1805.00034},
 primaryClass = {astro-ph.CO},
       adsurl = {https://ui.adsabs.harvard.edu/abs/2019A&A...621A..40E}
}

@ARTICLE{Markevitch2005,
       author = {{Markevitch}, M. and {Govoni}, F. and {Brunetti}, G. and {Jerius}, D.},
        title = "{Bow Shock and Radio Halo in the Merging Cluster A520}",
      journal = {\apj},
         year = 2005,
        month = jul,
       volume = {627},
       number = {2},
        pages = {733-738},
          doi = {10.1086/430695},
archivePrefix = {arXiv},
       eprint = {astro-ph/0412451},
 primaryClass = {astro-ph},
       adsurl = {https://ui.adsabs.harvard.edu/abs/2005ApJ...627..733M}
}

@ARTICLE{Ascasibar2006,
       author = {{Ascasibar}, Yago and {Markevitch}, Maxim},
        title = "{The Origin of Cold Fronts in the Cores of Relaxed Galaxy Clusters}",
      journal = {\apj},
         year = 2006,
        month = oct,
       volume = {650},
       number = {1},
        pages = {102-127},
          doi = {10.1086/506508},
archivePrefix = {arXiv},
       eprint = {astro-ph/0603246},
 primaryClass = {astro-ph},
       adsurl = {https://ui.adsabs.harvard.edu/abs/2006ApJ...650..102A}
}

@ARTICLE{McDonald2009,
       author = {{McDonald}, Michael and {Veilleux}, Sylvain},
        title = "{MMTF-H{\ensuremath{\alpha}} and HST-FUV Imaging of the Filamentary Complex in ABELL 1795}",
      journal = {\apjl},
         year = 2009,
        month = oct,
       volume = {703},
       number = {2},
        pages = {L172-L177},
          doi = {10.1088/0004-637X/703/2/L172},
archivePrefix = {arXiv},
       eprint = {0909.1554},
 primaryClass = {astro-ph.CO},
       adsurl = {https://ui.adsabs.harvard.edu/abs/2009ApJ...703L.172M}
}

@ARTICLE{Watson2023,
       author = {{Watson}, Courtney B. and {Blanton}, Elizabeth L. and {Randall}, Scott W. and {Sarazin}, Craig L. and {Sarkar}, Arnab and {ZuHone}, John A. and {Douglass}, E.~M.},
        title = "{CHANDRA X-Ray Observations of A119: Cold Fronts and a Shock in an Evolved Off-axis Merger}",
      journal = {\apj},
         year = 2023,
        month = oct,
       volume = {955},
       number = {2},
          eid = {103},
        pages = {103},
          doi = {10.3847/1538-4357/acee74},
archivePrefix = {arXiv},
       eprint = {2308.04367},
 primaryClass = {astro-ph.GA},
       adsurl = {https://ui.adsabs.harvard.edu/abs/2023ApJ...955..103W}
}

@ARTICLE{Paterno2013,
       author = {{Paterno-Mahler}, R. and {Blanton}, E.~L. and {Randall}, S.~W. and {Clarke}, T.~E.},
        title = "{Deep Chandra Observations of the Extended Gas Sloshing Spiral in A2029}",
      journal = {\apj},
         year = 2013,
        month = aug,
       volume = {773},
       number = {2},
          eid = {114},
        pages = {114},
          doi = {10.1088/0004-637X/773/2/114},
archivePrefix = {arXiv},
       eprint = {1306.3520},
 primaryClass = {astro-ph.CO},
       adsurl = {https://ui.adsabs.harvard.edu/abs/2013ApJ...773..114P}
}

@ARTICLE{Watson2025,
  author = {{Watson}, Courtney B. and
            {Blanton}, Elizabeth L. and
            {Randall}, Scott W. and
            {Clarke}, Tracy E. and
            {ZuHone}, John A.},
  title = "{Deep Chandra X-ray Observations of Abell 2029:
            the Merger History of a Relaxed, Strong Cool Core Cluster}",
  journal = {ApJ},
  year = {2025},
  note = {in press}
}

@ARTICLE{Rahaman2022,
       author = {{Rahaman}, Majidul and {Raja}, Ramij and {Datta}, Abhirup},
        title = "{On the detection of multiple shock fronts in A1914 using deep Chandra X-ray observations}",
      journal = {\mnras},
         year = 2022,
        month = feb,
       volume = {509},
       number = {4},
        pages = {5821-5835},
          doi = {10.1093/mnras/stab3115},
archivePrefix = {arXiv},
       eprint = {2110.12297},
 primaryClass = {astro-ph.HE},
       adsurl = {https://ui.adsabs.harvard.edu/abs/2022MNRAS.509.5821R}
}

@ARTICLE{Roediger2011,
       author = {{Roediger}, E. and {Br{\"u}ggen}, M. and {Simionescu}, A. and {B{\"o}hringer}, H. and {Churazov}, E. and {Forman}, W.~R.},
        title = "{Gas sloshing, cold front formation and metal redistribution: the Virgo cluster as a quantitative test case}",
      journal = {\mnras},
         year = 2011,
        month = may,
       volume = {413},
       number = {3},
        pages = {2057-2077},
          doi = {10.1111/j.1365-2966.2011.18279.x},
archivePrefix = {arXiv},
       eprint = {1007.4209},
 primaryClass = {astro-ph.CO},
       adsurl = {https://ui.adsabs.harvard.edu/abs/2011MNRAS.413.2057R}
}

@ARTICLE{Zhao2025,
       author = {{Zhao}, Hai-Sheng and {Li}, Cheng-Kui and {Wang}, Jin and {Zhang}, Juan and {Jia}, Shu-Mei and {Guan}, Ju and {Zhao}, Xiao-Fan and {Chen}, Yong and {Xu}, Jing-Jing and {Han}, Da-Wei and {Song}, Li-Ming and {Cui}, Wei-Wei},
        title = "{Data reduction and processing for the Follow-up X-ray Telescope onboard Einstein Probe}",
      journal = {Radiation Detection Technology and Methods},
         year = 2025,
        month = jun,
       volume = {9},
       number = {2},
        pages = {215-222},
          doi = {10.1007/s41605-025-00526-8},
       adsurl = {https://ui.adsabs.harvard.edu/abs/2025RDTM....9..215Z}
}

@ARTICLE{Willingale2013,
       author = {{Willingale}, R. and {Starling}, R.~L.~C. and {Beardmore}, A.~P. and {Tanvir}, N.~R. and {O'Brien}, P.~T.},
        title = "{Calibration of X-ray absorption in our Galaxy}",
      journal = {\mnras},
         year = 2013,
        month = may,
       volume = {431},
       number = {1},
        pages = {394-404},
          doi = {10.1093/mnras/stt175},
archivePrefix = {arXiv},
       eprint = {1303.0843},
 primaryClass = {astro-ph.HE},
       adsurl = {https://ui.adsabs.harvard.edu/abs/2013MNRAS.431..394W}
}

@ARTICLE{Yuan2025,
       author = {{Yuan}, Weimin and {Dai}, Lixin and {Feng}, Hua and {Jin}, Chichuan and {Jonker}, Peter and {Kuulkers}, Erik and {Liu}, Yuan and {Nandra}, Kirpal and {O'Brien}, Paul and {Piro}, Luigi and {Rau}, Arne and {Rea}, Nanda and {Sanders}, Jeremy and {Tao}, Lian and {Wang}, Junfeng and {Wu}, Xuefeng and {Zhang}, Bing and {Zhang}, Shuangnan and {Ai}, Shunke and {Buchner}, Johannes and {Bulbul}, Esra and {Chen}, Hechao and {Chen}, Minghua and {Chen}, Yong and {Chen}, Yu-Peng and {Coleiro}, Alexis and {Coti Zelati}, Francesco and {Dai}, Zigao and {Fan}, Xilong and {Fan}, Zhou and {Friedrich}, Susanne and {Gao}, He and {Ge}, Chong and {Ge}, Mingyu and {Geng}, Jinjun and {Ghirlanda}, Giancarlo and {Gianfagna}, Giulia and {Gou}, Lijun and {Guillot}, S{\'e}bastien and {Hou}, Xian and {Hu}, Jingwei and {Huang}, Yongfeng and {Ji}, Long and {Jia}, Shumei and {Komossa}, S. and {Kong}, Albert K.~H. and {Lan}, Lin and {Li}, An and {Li}, Ang and {Li}, Chengkui and {Li}, Dongyue and {Li}, Jian and {Li}, Zhaosheng and {Ling}, Zhixing and {Liu}, Ang and {Liu}, Jinzhong and {Liu}, Liangduan and {Liu}, Zhu and {Luo}, Jiawei and {Ma}, Ruican and {Maggi}, Pierre and {Maitra}, Chandreyee and {Marino}, Alessio and {Ng}, Stephen Chi-Yung and {Pan}, Haiwu and {Rukdee}, Surangkhana and {Soria}, Roberto and {Sun}, Hui and {Tam}, Pak-Hin Thomas and {Thakur}, Aishwarya Linesh and {Tian}, Hui and {Troja}, Eleonora and {Wang}, Wei and {Wang}, Xiangyu and {Wang}, Yanan and {Wei}, Junjie and {Wen}, Sixiang and {Wu}, Jianfeng and {Wu}, Ting and {Xiao}, Di and {Xu}, Dong and {Xu}, Renxin and {Xu}, Yanjun and {Xu}, Yu and {Yang}, Haonan and {You}, Bei and {Yu}, Heng and {Yu}, Yunwei and {Zhang}, Binbin and {Zhang}, Chen and {Zhang}, Guobao and {Zhang}, Liang and {Zhang}, Wenda and {Zhang}, Yu and {Zhou}, Ping and {Zou}, Zecheng},
        title = "{Science objectives of the Einstein Probe mission}",
      journal = {Science China Physics, Mechanics, and Astronomy},
         year = 2025,
        month = mar,
       volume = {68},
       number = {3},
          eid = {239501},
        pages = {239501},
          doi = {10.1007/s11433-024-2600-3},
archivePrefix = {arXiv},
       eprint = {2501.07362},
 primaryClass = {astro-ph.HE},
       adsurl = {https://ui.adsabs.harvard.edu/abs/2025SCPMA..6839501Y}
}

@ARTICLE{Churazov2003,
       author = {{Churazov}, E. and {Forman}, W. and {Jones}, C. and {B{\"o}hringer}, H.},
        title = "{XMM-Newton Observations of the Perseus Cluster. I. The Temperature and Surface Brightness Structure}",
      journal = {\apj},
         year = 2003,
        month = jun,
       volume = {590},
       number = {1},
        pages = {225-237},
          doi = {10.1086/374923},
archivePrefix = {arXiv},
       eprint = {astro-ph/0301482},
 primaryClass = {astro-ph},
       adsurl = {https://ui.adsabs.harvard.edu/abs/2003ApJ...590..225C}
}

@ARTICLE{Sanders2006,
       author = {{Sanders}, J.~S.},
        title = "{Contour binning: a new technique for spatially resolved X-ray spectroscopy applied to Cassiopeia A}",
      journal = {\mnras},
         year = 2006,
        month = sep,
       volume = {371},
       number = {2},
        pages = {829-842},
          doi = {10.1111/j.1365-2966.2006.10716.x},
archivePrefix = {arXiv},
       eprint = {astro-ph/0606528},
 primaryClass = {astro-ph},
       adsurl = {https://ui.adsabs.harvard.edu/abs/2006MNRAS.371..829S}
}

@INCOLLECTION{Yuan2022,
       author = {{Yuan}, Weimin and {Zhang}, Chen and {Chen}, Yong and {Ling}, Zhixing},
        title = "{The Einstein Probe Mission}",
    booktitle = {Handbook of X-ray and Gamma-ray Astrophysics},
         year = 2022,
       editor = {{Bambi}, Cosimo and {Sangangelo}, Andrea},
          eid = {86},
        publisher = {Springer},
        pages = {86},
          doi = {10.1007/978-981-16-4544-0_151-1},
       adsurl = {https://ui.adsabs.harvard.edu/abs/2022hxga.book...86Y}
}

@ARTICLE{sasaki2016,
       author = {{Sasaki}, Toru and {Matsushita}, Kyoko and {Sato}, Kosuke and {Okabe}, Nobuhiro},
        title = "{X-ray observations of a subhalo associated with the NGC 4839 group infalling toward the Coma cluster}",
      journal = {\pasj},
         year = 2016,
        month = oct,
       volume = {68},
       number = {5},
          eid = {85},
        pages = {85},
          doi = {10.1093/pasj/psw078},
archivePrefix = {arXiv},
       eprint = {1607.07554},
 primaryClass = {astro-ph.CO},
       adsurl = {https://ui.adsabs.harvard.edu/abs/2016PASJ...68...85S}
}

@ARTICLE{Ghirardini2019,
       author = {{Ghirardini}, V. and {Eckert}, D. and {Ettori}, S. and {Pointecouteau}, E. and {Molendi}, S. and {Gaspari}, M. and {Rossetti}, M. and {De Grandi}, S. and {Roncarelli}, M. and {Bourdin}, H. and {Mazzotta}, P. and {Rasia}, E. and {Vazza}, F.},
        title = "{Universal thermodynamic properties of the intracluster medium over two decades in radius in the X-COP sample}",
      journal = {\aap},
         year = 2019,
        month = jan,
       volume = {621},
          eid = {A41},
        pages = {A41},
          doi = {10.1051/0004-6361/201833325},
archivePrefix = {arXiv},
       eprint = {1805.00042},
 primaryClass = {astro-ph.CO},
       adsurl = {https://ui.adsabs.harvard.edu/abs/2019A&A...621A..41G}
}

@ARTICLE{Fabian2006,
       author = {{Fabian}, A.~C. and {Sanders}, J.~S. and {Taylor}, G.~B. and {Allen}, S.~W. and {Crawford}, C.~S. and {Johnstone}, R.~M. and {Iwasawa}, K.},
        title = "{A very deep Chandra observation of the Perseus cluster: shocks, ripples and conduction}",
      journal = {\mnras},
         year = 2006,
        month = feb,
       volume = {366},
       number = {2},
        pages = {417-428},
          doi = {10.1111/j.1365-2966.2005.09896.x},
archivePrefix = {arXiv},
       eprint = {astro-ph/0510476},
 primaryClass = {astro-ph},
       adsurl = {https://ui.adsabs.harvard.edu/abs/2006MNRAS.366..417F}
}

@ARTICLE{Kravtsov2012,
       author = {{Kravtsov}, Andrey V. and {Borgani}, Stefano},
        title = "{Formation of Galaxy Clusters}",
      journal = {\araa},
         year = 2012,
        month = sep,
       volume = {50},
        pages = {353-409},
          doi = {10.1146/annurev-astro-081811-125502},
archivePrefix = {arXiv},
       eprint = {1205.5556},
 primaryClass = {astro-ph.CO},
       adsurl = {https://ui.adsabs.harvard.edu/abs/2012ARA&A..50..353K}
}

@article{Zhang2025,
   title={Characteristics and Modeling of the In-flight Instrumental Background of FXT Onboard Einstein Probe},
   volume={25},
   ISSN={2397-6209},
   url={http://dx.doi.org/10.1088/1674-4527/ae05f8},
   DOI={10.1088/1674-4527/ae05f8},
   number={11},
   journal={Research in Astronomy and Astrophysics},
   publisher={IOP Publishing},
   author={Zhang, Juan and Chen, Yong and Jia, Shumei and Zhao, Haisheng and Cui, Weiwei and Chen, Tianxiang and Wang, Juan and Wang, Hao and Wang, Jin and Li, Chengkui and Zhao, Xiaofan and Guan, Ju and Han, Dawei and Xu, Jingjing and Song, Liming and Feng, Hua and Zhang, Shuangnan and Yuan, Weimin},
   year={2025},
   month=oct, pages={115019} }

@ARTICLE{Salom2003,
       author = {{Salom{\'e}}, P. and {Combes}, F.},
        title = "{Cold molecular gas in cooling flow clusters of galaxies}",
      journal = {\aap},
         year = 2003,
        month = dec,
       volume = {412},
        pages = {657-667},
          doi = {10.1051/0004-6361:20031438},
archivePrefix = {arXiv},
       eprint = {astro-ph/0309304},
 primaryClass = {astro-ph},
       adsurl = {https://ui.adsabs.harvard.edu/abs/2003A&A...412..657S}
}

@ARTICLE{Eckert2020,
       author = {{Eckert}, Dominique and {Finoguenov}, Alexis and {Ghirardini}, Vittorio and {Grandis}, Sebastian and {Kaefer}, Florian and {Sanders}, Jeremy and {Ramos-Ceja}, Miriam},
        title = "{Low-scatter galaxy cluster mass proxies for the eROSITA all-sky survey}",
      journal = {The Open Journal of Astrophysics},
         year = 2020,
        month = sep,
       volume = {3},
          eid = {12},
        pages = {12},
          doi = {10.21105/astro.2009.13944},
archivePrefix = {arXiv},
       eprint = {2009.03944},
 primaryClass = {astro-ph.CO},
       adsurl = {https://ui.adsabs.harvard.edu/abs/2020OJAp....3E..12E}
}

@ARTICLE{Mirakhor2023,
       author = {{Mirakhor}, M.~S. and {Walker}, S.~A. and {Runge}, J.},
        title = "{A deep dive: Chandra observations of the NGC 4839 group falling into the Coma cluster}",
      journal = {\mnras},
         year = 2023,
        month = jun,
       volume = {522},
       number = {2},
        pages = {2105-2114},
          doi = {10.1093/mnras/stad1088},
archivePrefix = {arXiv},
       eprint = {2304.05419},
 primaryClass = {astro-ph.GA},
       adsurl = {https://ui.adsabs.harvard.edu/abs/2023MNRAS.522.2105M}
}

@ARTICLE{Bourdin2008,
       author = {{Bourdin}, H. and {Mazzotta}, P.},
        title = "{Temperature structure of the intergalactic medium within seven nearby and bright clusters of galaxies observed with XMM-Newton}",
      journal = {\aap},
         year = 2008,
        month = feb,
       volume = {479},
       number = {2},
        pages = {307-320},
          doi = {10.1051/0004-6361:20065758},
archivePrefix = {arXiv},
       eprint = {0802.1866},
 primaryClass = {astro-ph},
       adsurl = {https://ui.adsabs.harvard.edu/abs/2008A&A...479..307B}
}

@ARTICLE{Sanders2016MNRAS,
       author = {{Sanders}, J.~S. and {Fabian}, A.~C. and {Taylor}, G.~B. and {Russell}, H.~R. and {Blundell}, K.~M. and {Canning}, R.~E.~A. and {Hlavacek-Larrondo}, J. and {Walker}, S.~A. and {Grimes}, C.~K.},
        title = "{A very deep Chandra view of metals, sloshing and feedback in the Centaurus cluster of galaxies}",
      journal = {\mnras},
         year = 2016,
        month = mar,
       volume = {457},
       number = {1},
        pages = {82-109},
          doi = {10.1093/mnras/stv2972},
archivePrefix = {arXiv},
       eprint = {1601.01489},
 primaryClass = {astro-ph.GA},
       adsurl = {https://ui.adsabs.harvard.edu/abs/2016MNRAS.457...82S}
}

@ARTICLE{WH2024,
       author = {{Wen}, Z.~L. and {Han}, J.~L.},
        title = "{A Catalog of 1.58 Million Clusters of Galaxies Identified from the DESI Legacy Imaging Surveys}",
      journal = {\apjs},
         year = 2024,
        month = jun,
       volume = {272},
       number = {2},
          eid = {39},
        pages = {39},
          doi = {10.3847/1538-4365/ad409d},
archivePrefix = {arXiv},
       eprint = {2404.02002},
 primaryClass = {astro-ph.CO},
       adsurl = {https://ui.adsabs.harvard.edu/abs/2024ApJS..272...39W}
}

@ARTICLE{WH2015,
       author = {{Wen}, Z.~L. and {Han}, J.~L.},
        title = "{Calibration of the Optical Mass Proxy for Clusters of Galaxies and an Update of the WHL12 Cluster Catalog}",
      journal = {\apj},
         year = 2015,
        month = jul,
       volume = {807},
       number = {2},
          eid = {178},
        pages = {178},
          doi = {10.1088/0004-637X/807/2/178},
archivePrefix = {arXiv},
       eprint = {1506.04503},
 primaryClass = {astro-ph.GA},
       adsurl = {https://ui.adsabs.harvard.edu/abs/2015ApJ...807..178W}
}

@ARTICLE{ZuHone2018,
       author = {{ZuHone}, J.~A. and {Kowalik}, K. and {{\"O}hman}, E. and {Lau}, E. and {Nagai}, D.},
        title = "{The Galaxy Cluster Merger Catalog: An Online Repository of Mock Observations from Simulated Galaxy Cluster Mergers}",
      journal = {\apjs},
         year = 2018,
        month = jan,
       volume = {234},
       number = {1},
          eid = {4},
        pages = {4},
          doi = {10.3847/1538-4365/aa99db},
archivePrefix = {arXiv},
       eprint = {1609.04121},
 primaryClass = {astro-ph.CO},
       adsurl = {https://ui.adsabs.harvard.edu/abs/2018ApJS..234....4Z}
}

@ARTICLE{ZuHone2011,
       author = {{ZuHone}, J.~A.},
        title = "{A Parameter Space Exploration of Galaxy Cluster Mergers. I. Gas Mixing and the Generation of Cluster Entropy}",
      journal = {\apj},
         year = 2011,
        month = feb,
       volume = {728},
       number = {1},
          eid = {54},
        pages = {54},
          doi = {10.1088/0004-637X/728/1/54},
archivePrefix = {arXiv},
       eprint = {1004.3820},
 primaryClass = {astro-ph.CO},
       adsurl = {https://ui.adsabs.harvard.edu/abs/2011ApJ...728...54Z}
}

@ARTICLE{Markevitch2002,
       author = {{Markevitch}, M. and {Gonzalez}, A.~H. and {David}, L. and {Vikhlinin}, A. and {Murray}, S. and {Forman}, W. and {Jones}, C. and {Tucker}, W.},
        title = "{A Textbook Example of a Bow Shock in the Merging Galaxy Cluster 1E 0657-56}",
      journal = {\apjl},
         year = 2002,
        month = mar,
       volume = {567},
       number = {1},
        pages = {L27-L31},
          doi = {10.1086/339619},
archivePrefix = {arXiv},
       eprint = {astro-ph/0110468},
 primaryClass = {astro-ph},
       adsurl = {https://ui.adsabs.harvard.edu/abs/2002ApJ...567L..27M}
}

@ARTICLE{Russell2010,
       author = {{Russell}, H.~R. and {Sanders}, J.~S. and {Fabian}, A.~C. and {Baum}, S.~A. and {Donahue}, M. and {Edge}, A.~C. and {McNamara}, B.~R. and {O'Dea}, C.~P.},
        title = "{Chandra observation of two shock fronts in the merging galaxy cluster Abell 2146}",
      journal = {\mnras},
         year = 2010,
        month = aug,
       volume = {406},
       number = {3},
        pages = {1721-1733},
          doi = {10.1111/j.1365-2966.2010.16822.x},
archivePrefix = {arXiv},
       eprint = {1004.1559},
 primaryClass = {astro-ph.CO},
       adsurl = {https://ui.adsabs.harvard.edu/abs/2010MNRAS.406.1721R}
}

@ARTICLE{Fabian1994,
       author = {{Fabian}, A.~C. and {Arnaud}, K.~A. and {Bautz}, M.~W. and {Tawara}, Y.},
        title = "{ASCA Observations of Cooling Flows in Clusters of Galaxies}",
      journal = {\apjl},
         year = 1994,
        month = nov,
       volume = {436},
        pages = {L63},
          doi = {10.1086/187633},
       adsurl = {https://ui.adsabs.harvard.edu/abs/1994ApJ...436L..63F}
}

@article{Sanderson2003,
   title={The Birmingham--CfA cluster scaling project -- I.Gas fraction and the M--TX relation},
   volume={340},
   ISSN={1365-2966},
   url={http://dx.doi.org/10.1046/j.1365-8711.2003.06401.x},
   DOI={10.1046/j.1365-8711.2003.06401.x},
   number={3},
   journal={Monthly Notices of the Royal Astronomical Society},
   publisher={Oxford University Press (OUP)},
   author={Sanderson, A. J. R. and Ponman, T. J. and Finoguenov, A. and Lloyd-Davies, E. J. and Markevitch, M.},
   year={2003},
   month=apr, pages={989–1010} }

@ARTICLE{DESI2026,
       author = {{DESI Collaboration} and {Abdul Karim}, M. and {Adame}, A.~G. and {Aguado}, D. and {Aguilar}, J. and {Ahlen}, S. and {Alam}, S. and {Aldering}, G. and {Alexander}, D.~M. and {Alfarsy}, R. and {Allen}, L. and {Allende Prieto}, C. and {Alves}, O. and {Anand}, A. and {Andrade}, U. and {Armengaud}, E. and {Avila}, S. and {Aviles}, A. and {Awan}, H. and {Bailey}, S. and {Baleato Lizancos}, A. and {Ballester}, O. and {Bault}, A. and {Bautista}, J. and {Bean}, R. and {Behera}, J. and {BenZvi}, S. and {Beraldo e Silva}, L. and {Bermejo-Climent}, J.~R. and {Beutler}, F. and {Bianchi}, D. and {Blake}, C. and {Blum}, R. and {Bolton}, A.~S. and {Bonici}, M. and {Brieden}, S. and {Brodzeller}, A. and {Brooks}, D. and {Buckley-Geer}, E. and {Burtin}, E. and {Bystr{\"o}m}, A. and {Canning}, R. and {Carnero Rosell}, A. and {Carr}, A. and {Carrilho}, P. and {Casas}, L. and {Castander}, F.~J. and {Cereskaite}, R. and {Cervantes-Cota}, J.~L. and {Chaussidon}, E. and {Chaves-Montero}, J. and {Chen}, S. and {Chen}, X. and {Circosta}, C. and {Claybaugh}, T. and {Cole}, S. and {Cooper}, A.~P. and {Cousinou}, M.-C. and {Cuceu}, A. and {Davis}, T.~M. and {Dawson}, K.~S. and {de Belsunce}, R. and {de la Cruz}, R. and {de la Macorra}, A. and {de Mattia}, A. and {Deiosso}, N. and {Della Costa}, J. and {Demina}, R. and {Demirbozan}, U. and {DeRose}, J. and {Dey}, A. and {Dey}, B. and {Ding}, J. and {Ding}, Z. and {Doel}, P. and {Douglass}, K. and {Dowicz}, M. and {Ebina}, H. and {Edelstein}, J. and {Eisenstein}, D.~J. and {Elbers}, W. and {Emas}, N. and {Escoffier}, S. and {Fagrelius}, P. and {Fan}, X. and {Fanning}, K. and {Favole}, G. and {Fawcett}, V.~A. and {Fern{\'a}ndez-Garc{\'\i}a}, E. and {Ferraro}, S. and {Findlay}, N. and {Font-Ribera}, A. and {Forero-Romero}, J.~E. and {Forero-S{\'a}nchez}, D. and {Frenk}, C.~S. and {G{\"a}nsicke}, B.~T. and {Galbany}, L. and {Garc{\'\i}a-Bellido}, J. and {Garcia-Quintero}, C. and {Garrison}, L.~H. and {Gazta{\~n}aga}, E. and {Gil-Mar{\'\i}n}, H. and {Gloudemans}, A. and {Gnedin}, O.~Y. and {Gontcho A Gontcho}, S. and {Gonzalez}, D. and {Gonzalez-Morales}, A.~X. and {Gonzalez-Perez}, V. and {Gordon}, C. and {Graur}, O. and {Green}, D. and {Gruen}, D. and {Gsponer}, R. and {Guandalin}, C. and {Gutierrez}, G. and {Guy}, J. and {Hahn}, C. and {Han}, J.~J. and {Han}, J. and {He}, S. and {Herrera-Alcantar}, H.~K. and {Heydenreich}, S. and {Honscheid}, K. and {Hou}, J. and {Howlett}, C. and {Huterer}, D. and {Ir{\v{s}}i{\v{c}}}, V. and {Ishak}, M. and {Jacques}, A. and {Jiang}, L. and {Jimenez}, J. and {Jing}, Y.~P. and {Joachimi}, B. and {Joudaki}, S. and {Joyce}, R. and {Jullo}, E. and {Juneau}, S. and {Kara{\c{c}}ayl{\i}}, N.~G. and {Karim}, T. and {Kehoe}, R. and {Kent}, S. and {Khederlarian}, A. and {Kirkby}, D. and {Kisner}, T. and {Kitaura}, F.-S. and {Kizhuprakkat}, N. and {Kong}, H. and {Koposov}, S.~E. and {Kremin}, A. and {Krolewski}, A. and {Lahav}, O. and {Lai}, Y. and {Lamman}, C. and {Lan}, T.-W. and {Landriau}, M. and {Lang}, D. and {Lange}, J.~U. and {Lasker}, J. and {Le Goff}, J.~M. and {Le Guillou}, L. and {Leauthaud}, A. and {Levi}, M.~E. and {Li}, S. and {Li}, T.~S. and {Liu}, W. and {Lodha}, K. and {Lokken}, M. and {Luo}, Y. and {Magneville}, C. and {Manera}, M. and {Manser}, C.~J. and {Margala}, D. and {Martini}, P. and {Maus}, M. and {McCullough}, J. and {McDonald}, P. and {Medina}, G.~E. and {Medina-Varela}, L. and {Meisner}, A. and {Mena-Fern{\'a}ndez}, J. and {Menegas}, A. and {Meneses-Rizo}, J. and {Mezcua}, M. and {Miquel}, R. and {Montero-Camacho}, P. and {Moon}, J. and {Moustakas}, J. and {Mu{\~n}oz-Guti{\'e}rrez}, A. and {Mu noz-Santos}, D. and {Myers}, A.~D. and {Myles}, J. and {Nadathur}, S. and {Najita}, J. and {Napolitano}, L. and {Newman}, J.~A. and {Nikakhtar}, F. and {Nikutta}, R. and {Niz}, G. and {Noriega}, H.~E. and {Nugent}, P.},
        title = "{Data Release 1 of the Dark Energy Spectroscopic Instrument}",
      journal = {\aj},
         year = 2026,
        month = may,
       volume = {171},
       number = {5},
          eid = {285},
        pages = {285},
          doi = {10.3847/1538-3881/ae4c43},
archivePrefix = {arXiv},
       eprint = {2503.14745},
 primaryClass = {astro-ph.CO},
       adsurl = {https://ui.adsabs.harvard.edu/abs/2026AJ....171..285D}
}

@ARTICLE{Dey2019,
       author = {{Dey}, Arjun and {Schlegel}, David J. and {Lang}, Dustin and {Blum}, Robert and {Burleigh}, Kaylan and {Fan}, Xiaohui and {Findlay}, Joseph R. and {Finkbeiner}, Doug and {Herrera}, David and {Juneau}, St{\'e}phanie and {Landriau}, Martin and {Levi}, Michael and {McGreer}, Ian and {Meisner}, Aaron and {Myers}, Adam D. and {Moustakas}, John and {Nugent}, Peter and {Patej}, Anna and {Schlafly}, Edward F. and {Walker}, Alistair R. and {Valdes}, Francisco and {Weaver}, Benjamin A. and {Y{\`e}che}, Christophe and {Zou}, Hu and {Zhou}, Xu and {Abareshi}, Behzad and {Abbott}, T.~M.~C. and {Abolfathi}, Bela and {Aguilera}, C. and {Alam}, Shadab and {Allen}, Lori and {Alvarez}, A. and {Annis}, James and {Ansarinejad}, Behzad and {Aubert}, Marie and {Beechert}, Jacqueline and {Bell}, Eric F. and {BenZvi}, Segev Y. and {Beutler}, Florian and {Bielby}, Richard M. and {Bolton}, Adam S. and {Brice{\~n}o}, C{\'e}sar and {Buckley-Geer}, Elizabeth J. and {Butler}, Karen and {Calamida}, Annalisa and {Carlberg}, Raymond G. and {Carter}, Paul and {Casas}, Ricard and {Castander}, Francisco J. and {Choi}, Yumi and {Comparat}, Johan and {Cukanovaite}, Elena and {Delubac}, Timoth{\'e}e and {DeVries}, Kaitlin and {Dey}, Sharmila and {Dhungana}, Govinda and {Dickinson}, Mark and {Ding}, Zhejie and {Donaldson}, John B. and {Duan}, Yutong and {Duckworth}, Christopher J. and {Eftekharzadeh}, Sarah and {Eisenstein}, Daniel J. and {Etourneau}, Thomas and {Fagrelius}, Parker A. and {Farihi}, Jay and {Fitzpatrick}, Mike and {Font-Ribera}, Andreu and {Fulmer}, Leah and {G{\"a}nsicke}, Boris T. and {Gaztanaga}, Enrique and {George}, Koshy and {Gerdes}, David W. and {Gontcho}, Satya Gontcho A. and {Gorgoni}, Claudio and {Green}, Gregory and {Guy}, Julien and {Harmer}, Diane and {Hernandez}, M. and {Honscheid}, Klaus and {Huang}, Lijuan Wendy and {James}, David J. and {Jannuzi}, Buell T. and {Jiang}, Linhua and {Joyce}, Richard and {Karcher}, Armin and {Karkar}, Sonia and {Kehoe}, Robert and {Kneib}, Jean-Paul and {Kueter-Young}, Andrea and {Lan}, Ting-Wen and {Lauer}, Tod R. and {Le Guillou}, Laurent and {Le Van Suu}, Auguste and {Lee}, Jae Hyeon and {Lesser}, Michael and {Perreault Levasseur}, Laurence and {Li}, Ting S. and {Mann}, Justin L. and {Marshall}, Robert and {Mart{\'\i}nez-V{\'a}zquez}, C.~E. and {Martini}, Paul and {du Mas des Bourboux}, H{\'e}lion and {McManus}, Sean and {Meier}, Tobias Gabriel and {M{\'e}nard}, Brice and {Metcalfe}, Nigel and {Mu{\~n}oz-Guti{\'e}rrez}, Andrea and {Najita}, Joan and {Napier}, Kevin and {Narayan}, Gautham and {Newman}, Jeffrey A. and {Nie}, Jundan and {Nord}, Brian and {Norman}, Dara J. and {Olsen}, Knut A.~G. and {Paat}, Anthony and {Palanque-Delabrouille}, Nathalie and {Peng}, Xiyan and {Poppett}, Claire L. and {Poremba}, Megan R. and {Prakash}, Abhishek and {Rabinowitz}, David and {Raichoor}, Anand and {Rezaie}, Mehdi and {Robertson}, A.~N. and {Roe}, Natalie A. and {Ross}, Ashley J. and {Ross}, Nicholas P. and {Rudnick}, Gregory and {Safonova}, Sasha and {Saha}, Abhijit and {S{\'a}nchez}, F. Javier and {Savary}, Elodie and {Schweiker}, Heidi and {Scott}, Adam and {Seo}, Hee-Jong and {Shan}, Huanyuan and {Silva}, David R. and {Slepian}, Zachary and {Soto}, Christian and {Sprayberry}, David and {Staten}, Ryan and {Stillman}, Coley M. and {Stupak}, Robert J. and {Summers}, David L. and {Sien Tie}, Suk and {Tirado}, H. and {Vargas-Maga{\~n}a}, Mariana and {Vivas}, A. Katherina and {Wechsler}, Risa H. and {Williams}, Doug and {Yang}, Jinyi and {Yang}, Qian and {Yapici}, Tolga and {Zaritsky}, Dennis and {Zenteno}, A. and {Zhang}, Kai and {Zhang}, Tianmeng and {Zhou}, Rongpu and {Zhou}, Zhimin},
        title = "{Overview of the DESI Legacy Imaging Surveys}",
      journal = {\aj},
         year = 2019,
        month = may,
       volume = {157},
       number = {5},
          eid = {168},
        pages = {168},
          doi = {10.3847/1538-3881/ab089d},
archivePrefix = {arXiv},
       eprint = {1804.08657},
 primaryClass = {astro-ph.IM},
       adsurl = {https://ui.adsabs.harvard.edu/abs/2019AJ....157..168D}
}

@ARTICLE{Abdurro2022,
       author = {{Abdurro'uf} and {Accetta}, Katherine and {Aerts}, Conny and {Silva Aguirre}, V{\'\i}ctor and {Ahumada}, Romina and {Ajgaonkar}, Nikhil and {Filiz Ak}, N. and {Alam}, Shadab and {Allende Prieto}, Carlos and {Almeida}, Andr{\'e}s and {Anders}, Friedrich and {Anderson}, Scott F. and {Andrews}, Brett H. and {Anguiano}, Borja and {Aquino-Ort{\'\i}z}, Erik and {Arag{\'o}n-Salamanca}, Alfonso and {Argudo-Fern{\'a}ndez}, Maria and {Ata}, Metin and {Aubert}, Marie and {Avila-Reese}, Vladimir and {Badenes}, Carles and {Barb{\'a}}, Rodolfo H. and {Barger}, Kat and {Barrera-Ballesteros}, Jorge K. and {Beaton}, Rachael L. and {Beers}, Timothy C. and {Belfiore}, Francesco and {Bender}, Chad F. and {Bernardi}, Mariangela and {Bershady}, Matthew A. and {Beutler}, Florian and {Bidin}, Christian Moni and {Bird}, Jonathan C. and {Bizyaev}, Dmitry and {Blanc}, Guillermo A. and {Blanton}, Michael R. and {Boardman}, Nicholas Fraser and {Bolton}, Adam S. and {Boquien}, M{\'e}d{\'e}ric and {Borissova}, Jura and {Bovy}, Jo and {Brandt}, W.~N. and {Brown}, Jordan and {Brownstein}, Joel R. and {Brusa}, Marcella and {Buchner}, Johannes and {Bundy}, Kevin and {Burchett}, Joseph N. and {Bureau}, Martin and {Burgasser}, Adam and {Cabang}, Tuesday K. and {Campbell}, Stephanie and {Cappellari}, Michele and {Carlberg}, Joleen K. and {Wanderley}, F{\'a}bio Carneiro and {Carrera}, Ricardo and {Cash}, Jennifer and {Chen}, Yan-Ping and {Chen}, Wei-Huai and {Cherinka}, Brian and {Chiappini}, Cristina and {Choi}, Peter Doohyun and {Chojnowski}, S. Drew and {Chung}, Haeun and {Clerc}, Nicolas and {Cohen}, Roger E. and {Comerford}, Julia M. and {Comparat}, Johan and {da Costa}, Luiz and {Covey}, Kevin and {Crane}, Jeffrey D. and {Cruz-Gonzalez}, Irene and {Culhane}, Connor and {Cunha}, Katia and {Dai}, Y. Sophia and {Damke}, Guillermo and {Darling}, Jeremy and {Davidson}, Jr., James W. and {Davies}, Roger and {Dawson}, Kyle and {De Lee}, Nathan and {Diamond-Stanic}, Aleksandar M. and {Cano-D{\'\i}az}, Mariana and {S{\'a}nchez}, Helena Dom{\'\i}nguez and {Donor}, John and {Duckworth}, Chris and {Dwelly}, Tom and {Eisenstein}, Daniel J. and {Elsworth}, Yvonne P. and {Emsellem}, Eric and {Eracleous}, Mike and {Escoffier}, Stephanie and {Fan}, Xiaohui and {Farr}, Emily and {Feng}, Shuai and {Fern{\'a}ndez-Trincado}, Jos{\'e} G. and {Feuillet}, Diane and {Filipp}, Andreas and {Fillingham}, Sean P. and {Frinchaboy}, Peter M. and {Fromenteau}, Sebastien and {Galbany}, Llu{\'\i}s and {Garc{\'\i}a}, Rafael A. and {Garc{\'\i}a-Hern{\'a}ndez}, D.~A. and {Ge}, Junqiang and {Geisler}, Doug and {Gelfand}, Joseph and {G{\'e}ron}, Tobias and {Gibson}, Benjamin J. and {Goddy}, Julian and {Godoy-Rivera}, Diego and {Grabowski}, Kathleen and {Green}, Paul J. and {Greener}, Michael and {Grier}, Catherine J. and {Griffith}, Emily and {Guo}, Hong and {Guy}, Julien and {Hadjara}, Massinissa and {Harding}, Paul and {Hasselquist}, Sten and {Hayes}, Christian R. and {Hearty}, Fred and {Hern{\'a}ndez}, Jes{\'u}s and {Hill}, Lewis and {Hogg}, David W. and {Holtzman}, Jon A. and {Horta}, Danny and {Hsieh}, Bau-Ching and {Hsu}, Chin-Hao and {Hsu}, Yun-Hsin and {Huber}, Daniel and {Huertas-Company}, Marc and {Hutchinson}, Brian and {Hwang}, Ho Seong and {Ibarra-Medel}, H{\'e}ctor J. and {Chitham}, Jacob Ider and {Ilha}, Gabriele S. and {Imig}, Julie and {Jaekle}, Will and {Jayasinghe}, Tharindu and {Ji}, Xihan and {Johnson}, Jennifer A. and {Jones}, Amy and {J{\"o}nsson}, Henrik and {Katkov}, Ivan and {Khalatyan}, Dr., Arman and {Kinemuchi}, Karen and {Kisku}, Shobhit and {Knapen}, Johan H. and {Kneib}, Jean-Paul and {Kollmeier}, Juna A. and {Kong}, Miranda and {Kounkel}, Marina and {Kreckel}, Kathryn and {Krishnarao}, Dhanesh and {Lacerna}, Ivan and {Lane}, Richard R. and {Langgin}, Rachel and {Lavender}, Ramon and {Law}, David R. and {Lazarz}, Daniel and {Leung}, Henry W. and {Leung}, Ho-Hin and {Lewis}, Hannah M. and {Li}, Cheng and {Li}, Ran and {Lian}, Jianhui and {Liang}, Fu-Heng and {Lin}, Lihwai and {Lin}, Yen-Ting and {Lin}, Sicheng and {Lintott}, Chris and {Long}, Dan and {Longa-Pe{\~n}a}, Pen{\'e}lope and {L{\'o}pez-Cob{\'a}}, Carlos and {Lu}, Shengdong and {Lundgren}, Britt F. and {Luo}, Yuanze and {Mackereth}, J. Ted and {de la Macorra}, Axel and {Mahadevan}, Suvrath and {Majewski}, Steven R. and {Manchado}, Arturo and {Mandeville}, Travis and {Maraston}, Claudia and {Margalef-Bentabol}, Berta and {Masseron}, Thomas and {Masters}, Karen L. and {Mathur}, Savita and {McDermid}, Richard M. and {Mckay}, Myles and {Merloni}, Andrea and {Merrifield}, Michael and {Meszaros}, Szabolcs and {Miglio}, Andrea and {Di Mille}, Francesco and {Minniti}, Dante and {Minsley}, Rebecca and {Monachesi}, Antonela},
        title = "{The Seventeenth Data Release of the Sloan Digital Sky Surveys: Complete Release of MaNGA, MaStar, and APOGEE-2 Data}",
      journal = {\apjs},
         year = 2022,
        month = apr,
       volume = {259},
       number = {2},
          eid = {35},
        pages = {35},
          doi = {10.3847/1538-4365/ac4414},
archivePrefix = {arXiv},
       eprint = {2112.02026},
 primaryClass = {astro-ph.GA},
       adsurl = {https://ui.adsabs.harvard.edu/abs/2022ApJS..259...35A}
}
\bibliographystyle{aa}

\end{document}